\documentclass[aps, twocolumn]{revtex4-1}
\usepackage{bm}
\usepackage[margin=1in]{geometry}
\usepackage[utf8]{inputenc}
\usepackage{amsfonts}
\usepackage{amssymb}
\usepackage{graphicx}
\usepackage{amsmath}
\usepackage{ragged2e}
\usepackage{empheq}
\usepackage{tikz}
\usetikzlibrary{arrows,positioning,shapes,backgrounds,fit}  
\usetikzlibrary{tikzmark,calc}
\usepackage{caption}
\usepackage{float}
\usepackage[normalem]{ulem}

\begin{document}

\begin{abstract}
We use extreme value statistics to study the dynamics of coarsening in
aggregation-fragmentation models which form condensates in the steady state. The dynamics is dominated by the formation of local condensates on a coarsening length scale which grows in time in both the zero range process and conserved mass aggregation model. The local condensate mass distribution exhibits scaling, which implies anomalously large fluctuations, with mean and standard deviation both proportional to the coarsening length. Remarkably, the state of the system during
coarsening is governed not by the steady state, but rather a pre-asymptotic state in which the condensate mass fluctuates strongly.
\end{abstract}

\title{Coarsening, condensates and extremes in aggregation-fragmentation models}

\author{Chandrashekar Iyer$^{1,2}$, Arghya Das$^1$, Mustansir Barma$^1$}
\affiliation{$^1$TIFR Centre for Interdisciplinary Sciences, Tata Institute of Fundamental Research, Gopanpally, Hyderabad, 500046, India\\
$^2$UM-DAE Centre for Excellence in Basic Sciences, Mumbai 400098, India}

\maketitle

\section{\label{sec:level1}Introduction}

When a system with a propensity to order is prepared in a far-from-equilibrium disordered state, it takes a long time for order to get established. The process by which this happens, namely phase ordering or coarsening, is characterized by well-defined laws of broad validity \cite{Bray1994}. These laws apply in a wide variety of systems, ranging from magnets and alloys on the one hand  \cite{Bray1994, Puri-Wadhavan}, to non-equilibrium driven systems with well-ordered or fluctuation-dominated ordered states on the other \cite{Shauri2016,Dibyendu2001}. In all these systems there is an overarching similarity: order sets in over a coarsening length scale which grows with time.  Physically, microscopic degrees of freedom with separations smaller than this scale are strongly correlated, while for larger separations correlations are weak.

A natural expectation is that up to this scale, the state of the coarsening system should resemble the ordered steady state. Indeed this sort of steady state correspondence is a central tenet of coarsening theory and holds in the coarsening regime in the systems mentioned above. Surprisingly though, as we show below, it fails in a qualitative way in an important class of systems which evolve through aggregation-fragmentation dynamics.
This unusual failure is linked to the onset of extremely large fluctuations. These have simple scaling properties in the full coarsening regime, distinctly different from those in steady state; ultimately they can be traced to the occurrence of a pre-asymptotic state with large fluctuations. The result is robust with respect to different types of dynamics and holds for systems with both equilibrium and non-equilibrium steady states.

In this Article, we study coarsening in a class of mass transport models in which the kinetic moves involve diffusion, aggregation and fragmentation.  Such generic models have been used to caricature a broad range of systems including granular systems \cite{Eggers1999, Devaraj2004}, gelation \cite{Ziff-JSP1980}, transport in organelle in the cell \cite{Himani2013}, traffic flow \cite{Krug-Ferrari, Evans-Traffic}, and wealth accumulation \cite{Yakovenko2009}. An interesting feature that appears in several such models is the formation of a \emph{condensate}: in the steady state, a finite fraction of the total mass in the system resides on one or few condensate site(s) \cite{Evans2005, Evans-Zia-2006, Evans-Satya-2008, Satya-CMAM-1998, Evans-PFSS-2006}. Our study reveals that in an infinite system, there is indeed a characteristic length $\mathcal{L}(t)$ which grows indefinitely with time and governs coarsening \cite{definition}. However, the locally ordered state of the system on length scales up to and including $\mathcal{L}(t)$ is very different in character from the ordered steady state reached at large times in a finite system. This is demonstrated below for two mass transport models, \textcolor{black}{the Zero Range Process (ZRP) and the Conserved Mass Aggregation Model (CMAM)}, which differ in allowed kinetic moves, but share the common feature of condensate formation. A key point is that there are extremely large fluctuations of the mass contained in different regions of size $\mathcal{L}(t)$; the mean and standard deviation are both proportional to $\mathcal{L}(t)$, quite unlike the steady state.

The statistics of extreme values \cite{Satya2020} gives direct information about the condensate mass, and extreme value distributions (EVDs) have been used successfully to study the condensate mass in the steady state of the zero range process \cite{Satya-Zia-2005, Evans-Satya-2008}. Here we show that EVDs are extremely useful probes of condensate formation during coarsening as well. In the initial completely disordered state, masses at different sites are uncorrelated, and the EVD follows the well-known Gumbel form \cite{Satya2020}. On finally reaching the steady state at very large times in a finite system, overall mass conservation induces correlation and the EVD is quite different \cite{Satya-Zia-2005}. By monitoring the time evolution of the EVD in a region of size $\mathcal{L}(t)$, we show that it assumes a simple scaling form in the coarsening regime. However this form is in conflict with the result in steady state. \textcolor{black}{By studying the evolution of the global maximum mass in both models, we show the occurrence of a pre-asymptotic regime which governs the state in the coarsening system.}

\textcolor{black}{The article is structured as follows: In Section II, we introduce the two models and review their steady states. In Section III, we show numerical results for the correlation functions during coarsening and give evidence for a growing length scale $\mathcal{L}(t)$. Using the knowledge of $\mathcal{L}(t)$ above, in Section IV we obtain the probability distribution of the maximum mass in subsystems of size $\mathcal{L}(t)$, and demonstrate new scaling laws emerging during coarsening. In the next section, we investigate the time evolution of the global maximum mass in both models, and find corroboration of the anomalous $\mathcal{O}(\mathcal{L}(t))$ fluctuations from Section IV earlier. We also discuss the mechanism for these anomalous fluctuations and show that they are linked to fluctuations in a pre-asymptotic state. In Section VI, we explore the mapping to exclusion models and study the correlations during coarsening. Finally we conclude in Section VII.}

\section{\label{sec:level2}Models: Definitions and Steady States}

\begin{figure}[H]

    \includegraphics[width=3.8cm, height=4cm]{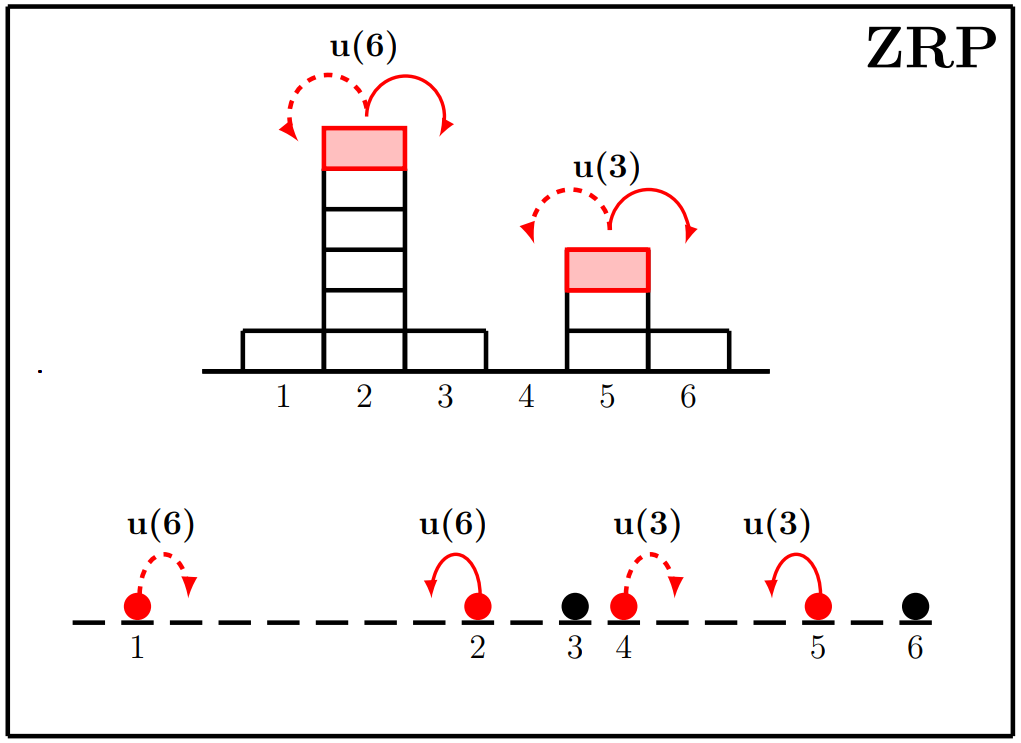}
    \includegraphics[width=3.8cm, height=4cm]{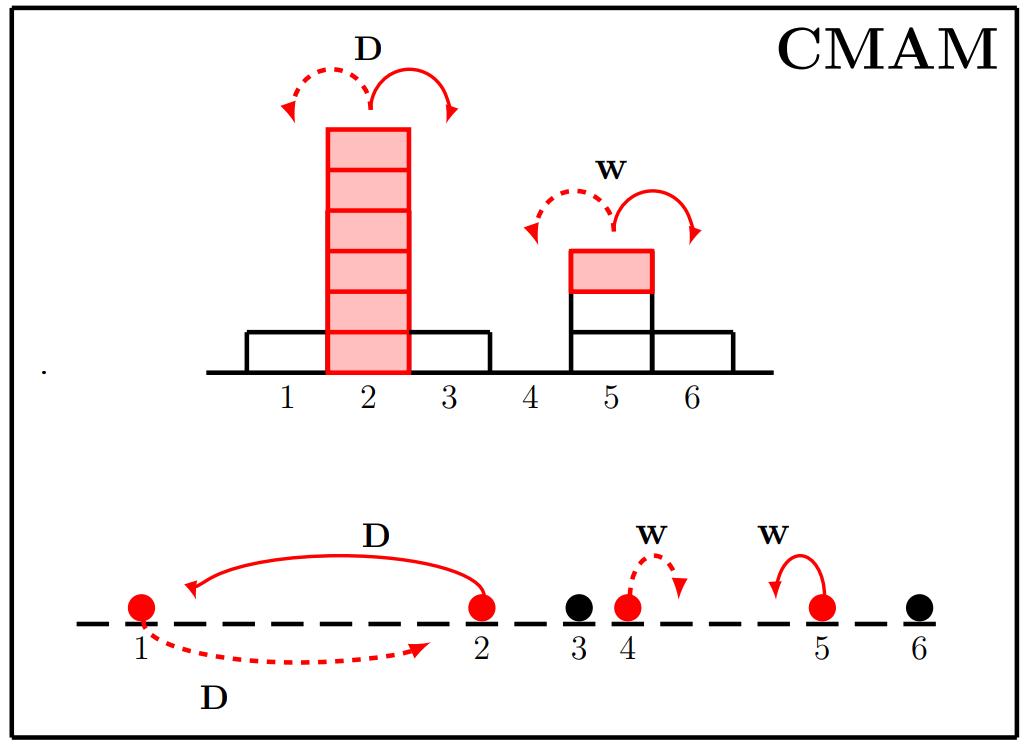}
\caption{\it \raggedright Illustrating ZRP and CMAM moves in the left and right panels respectively. The mapping to exclusion models is depicted in the lower part of each panel.}\label{schematic}
\end{figure}

We work with two well-studied models, namely the zero range process (ZRP) \cite{Spitzer,Evans2005} and the conserved mass aggregation model (CMAM) \cite{Satya-CMAM-1998, Rajesh-Satya-2001} on a 1D lattice ring with $L$ sites. Site $i$ contains $m_i$ particles, each with unit mass, with $m_i \ge 0$. The allowed moves conserve the total mass $M$. The system is specified by the overall density $\rho=M/L$ and the transition rates in Eqs. \eqref{ZRP} and \eqref{CMAM}, illustrated in Fig. \ref{schematic}.

The stochastic move in the asymmetric ZRP consists of chipping, namely the transfer of a single particle from an occupied site to its right neighbour, at a rate which depends on the occupancy, {\it viz.} $u(m_i)= 1+\frac{b}{m_i}$:
\vspace*{-0.5em}
\begin{equation}
(m_{i}, m_{i+1}) \xrightarrow{~u(m_i)~} (m_{i} -1, m_{i+1} +1). \label{ZRP}
\end{equation}
In the CMAM, in addition to chipping at rate $w$, the entire mass at site $i$ may be transferred to one of its neighbouring sites $i\pm 1$ at rate $D$:
\begin{eqnarray}
(m_{i}, m_{i \pm 1}) &\xrightarrow{w}& (m_{i} -1, m_{i \pm 1} +1), \nonumber\\
(m_{i}, m_{i \pm 1}) &\xrightarrow{D}& (0, m_{i \pm 1} + m_{i}). \label{CMAM}
\end{eqnarray}
In both models, the transferred mass coalesces with the mass already present on that site. \textcolor{black}{Below we discuss some properties of the two models in the steady state.}

\subsection{Steady State}

\subsubsection{ZRP}

At large times $\sim L^z$ with $z=2$, the ZRP with asymmetric hopping reaches a nonequilibrium steady state carrying a finite current \cite{Godreche2003, Grossinsky2003}. With symmetric hopping one finds $z=3$ \cite{Grossinsky2003}; in this case, detailed balance holds and we have an equilibrium state.
The steady state measure is identical in both cases \cite{Spitzer, Evans2005}.
For $b > 2$ there is a phase transition at density $\rho > \rho_c = 1/(b - 2)$ \cite{Evans2005}. The probability $p(m)$ that the mass at a given site is $m$ falls exponentially for large $m$ if $\rho < \rho_c$, and as $m^{-b}$ at $\rho = \rho_c$. For $\rho > \rho_c$, a condensate with mean mass $(\rho - \rho_c)L$ forms at one site and coexists with a background critical fluid of density $\rho_c$. Since the condensate mass $M_{\rm cond}$ is the largest in the system, its probability distribution can be found as an EVD \cite{Evans-Satya-2008, Satya2020, Barkai2020}. The result reflects correlations between $\lbrace m_i\rbrace$ arising from the conservation of total mass \cite{Evans-Zia-2006}:
\begin{equation}
P_{\rm ss}(M_{\rm cond})= \frac{1}{L^{\delta}} V_{\delta}\left(\frac{M_{\rm cond} - (\rho-\rho_c)L}{L^{\delta}} \right),
\end{equation}
where $\delta = 1/(b - 1)$ for $2<b<3$ and $\delta=1/2$ for $b>3$. In the latter case the scaling function is a Gaussian, whereas for $2<b<3$, $V_{\delta}(y)$ is highly non-Gaussian \cite{Evans-Zia-2006,Evans-Satya-2008}. The variance of the condensate mass $\sigma_{\rm cond}^2=\langle M_{\rm cond}^2 \rangle - \langle M_{\rm cond} \rangle^2 \sim O(L^{2/(b-1)})$ for $2<b<3$ and $O(L)$ for $b>3$ \cite{Evans-Zia-2006,Evans-Satya-2008}.

\subsubsection{CMAM}

In the CMAM with symmetric hopping, detailed balance does not hold and the steady state is far from equilibrium. Although the steady state measure is not known exactly, two-point correlations are known to factorise \cite{Rajesh-Satya-2001}. The system undergoes a phase transition at density $\rho_c =\sqrt{1+w/D}\,\, –\,\, 1$. For $\rho < \rho_c$, the probability $P(m)$ falls exponentially for large $m$, and as $m^{-5/2}$ for $\rho = \rho_c$. For $\rho>\rho_c$, a condensate with an average of $(\rho - \rho_c)L$ particles forms \cite{Rajesh-Satya-2001}. Unlike the ZRP, the condensate in the CMAM is mobile and diffuses in space. 

\begin{figure}[H]
 \begin{center}
 \includegraphics[scale=0.34]{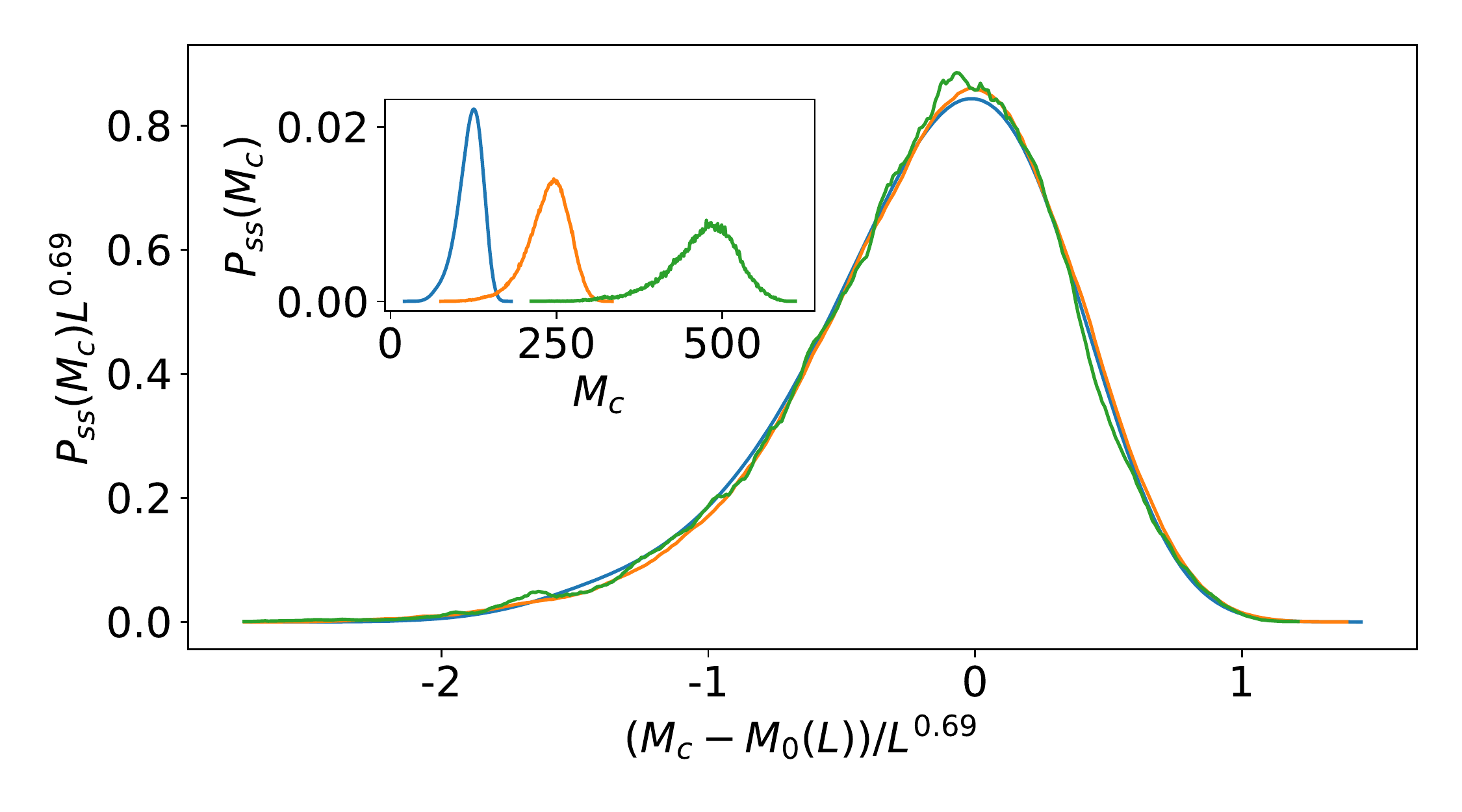}
 \caption{\it \raggedright Scaled distribution of the condensate mass in CMAM in steady state. $M_{0}(L)=$ $126,247,488$ for $L=200,400,800$ respectively. Inset: Unscaled distributions.}\label{pmmax-cmam}
\end{center}
\end{figure}

The EVD in steady state $P_{ss}(M_{\rm cond})$ is not known analytically. \textcolor{black}{From our numerical study we find that it is a scaling function of $\left(M_{\text {cond }}-M_{0}(L)\right) / L^{\beta}$ with $\beta \simeq 0.69$, implying that the mean and standard deviation grow as $L^{\beta}$. The scaling function has a non-Gaussian form (Fig. \ref{pmmax-cmam}). Here $M_{0}(L) \approx\left(\rho-\rho_{c}\right) L$ is the mode of the distribution.}   Note that there is no condensate formation in the CMAM with asymmetric hopping \cite{ACMAM}.

In the rest of the paper we have used $b=3.5$ in ZRP and $w=D=1$ in CMAM.

\section{Correlations during Coarsening}

 An important manifestation of the growing length scale $\mathcal{L}(t)$ in a coarsening system with density $\rho>\rho_c$ is that the two-point correlation function $G(r,t)\equiv \langle m_i(t)\,m_{i+r}(t)\rangle - \rho^2$ is a function of the scaled separation \cite{Bray1994},
\begin{equation}
G(r,t) = g(r/\mathcal{L}(t)),~ \mathcal{L}(t) \propto t^{1/z}\label{corr-scale}
\end{equation}

\begin{figure}[h]
    \includegraphics[width=7.4cm, height=4.5cm]{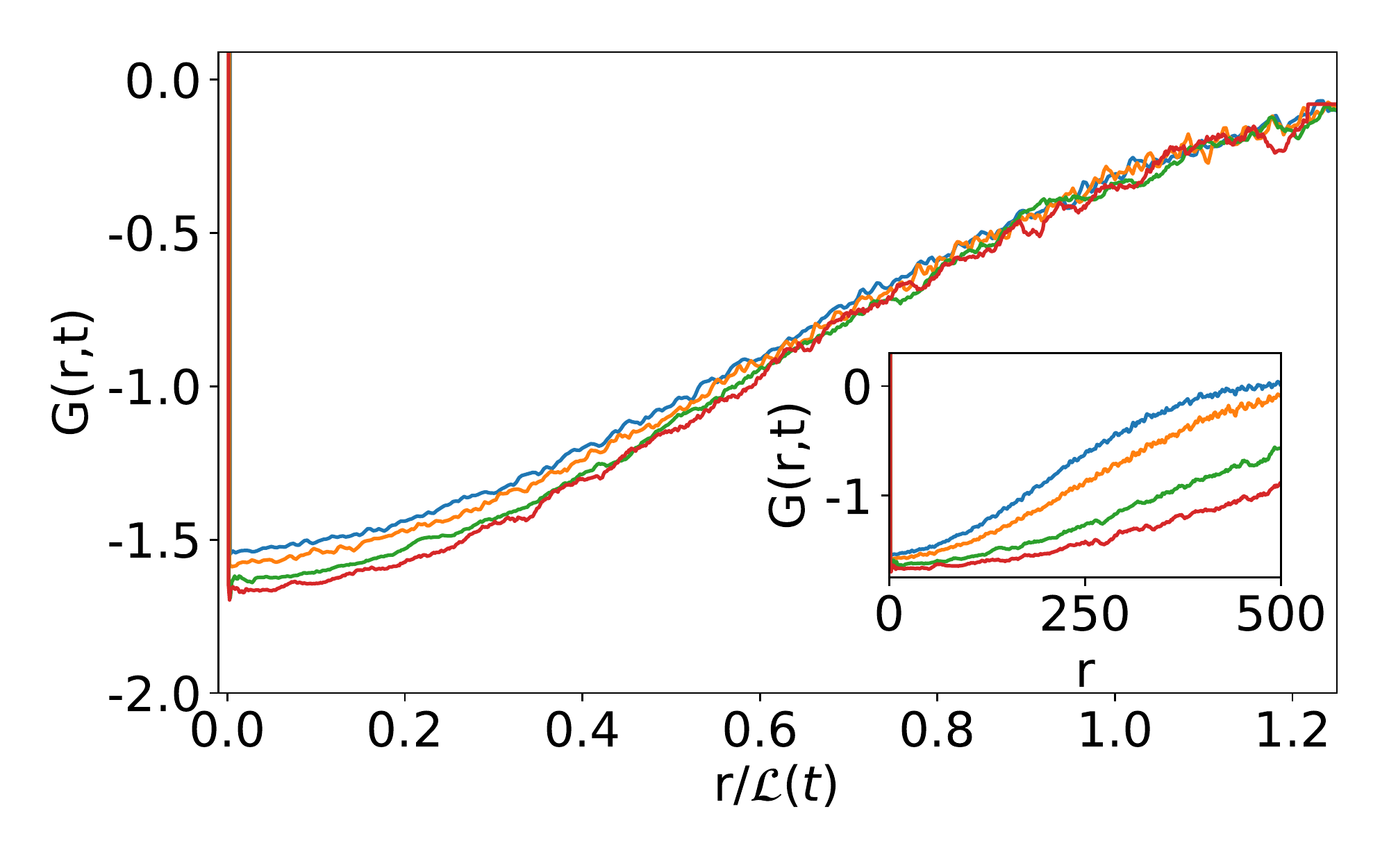}\\ 
    \includegraphics[width=8.3cm, height=4.5cm]{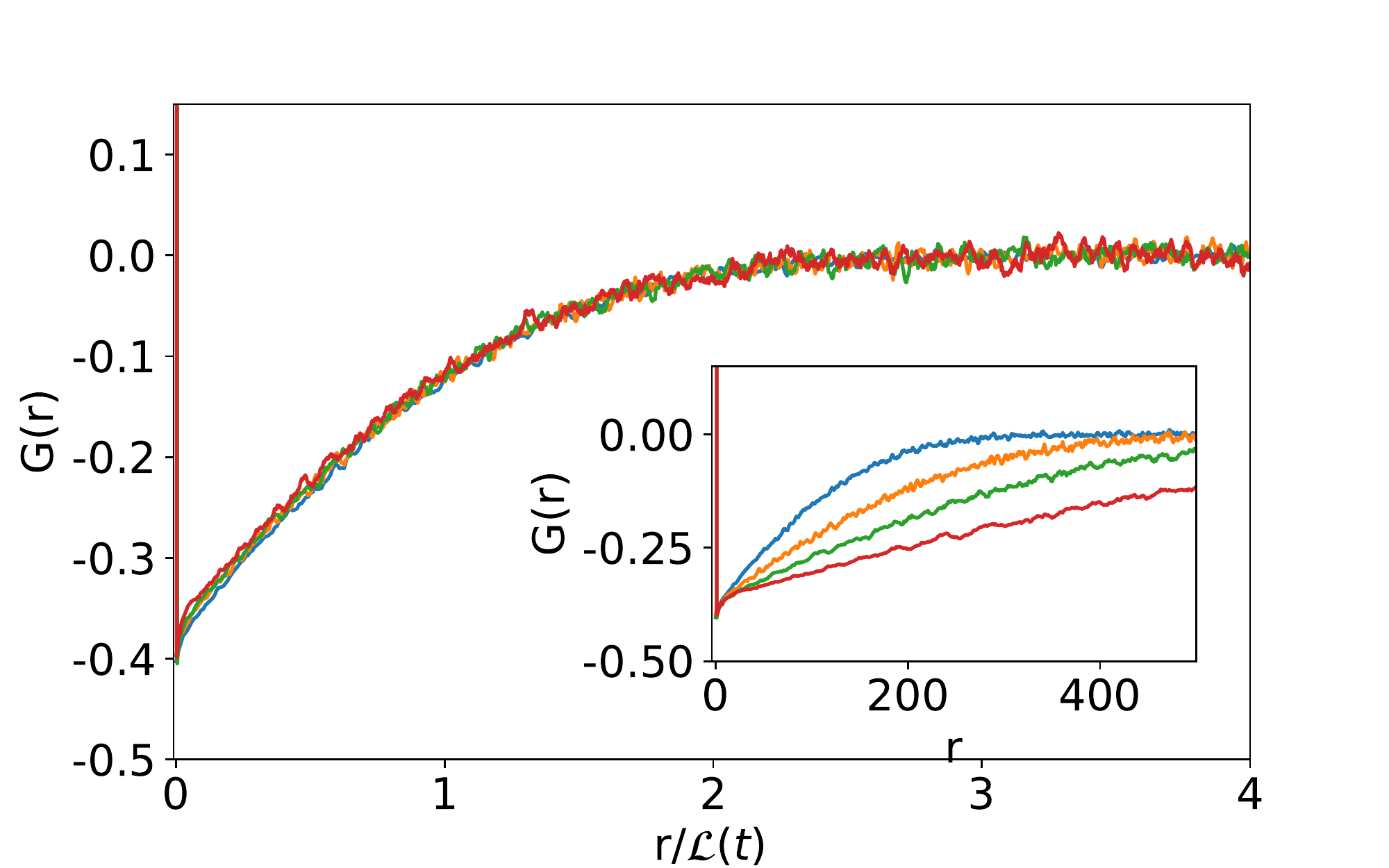}
\caption{\it \raggedright During coarsening, $G(r,t)$ is a function of the scaled separation $r/\mathcal{L}(t)$ with $\mathcal{L}(t)=t^{1/2}$. Top panel: asymmetric ZRP ($b=3.5,\rho_c=2/3,\rho=2$), with $\mathcal{L}(t)= 256,400,613,800$; Bottom panel: CMAM ($w=1,D=1,  \rho_c=\sqrt{2}-1,\rho=1$) with $\mathcal{L}(t)=120, 200, 300,500$. Insets: Unscaled plots. The system size $L=10,000$ for both models.}\label{correlation}
\end{figure}

Fig. \ref{correlation} shows scaling plots from numerical simulations of the asymmetric ZRP and CMAM using $\mathcal{L}(t)=\sqrt{t}$.
The scaling form Eq. \eqref{corr-scale} holds for both models, with $z=2$. This value agrees with the earlier determination of $z$ for the asymmetric ZRP \cite{Godreche2003, Grossinsky2003, Evans2005}. 

For $r=0$, $G(0,t)$ is large and positive $\sim O(\mathcal{L}(t)^2)$ whereas for $r\ne 0$, $G(r,t)$ is negative. This implies that the differences $\delta m_i = (m_i - \rho)$ and $\delta m_{i+r} = (m_{i+r} - \rho)$ are anti-correlated, which is consistent with a simple picture: In a region of size $\mathcal{L}(t)$, local condensates which hold a significant fraction of the mass ($\sim \mathcal{L}(t)$) within the region can form at one site (or few sites). At such sites, $\delta m_i$ is large and positive. However, $\delta m_i$ is typically negative on the remaining sites which hold the `coarsening critical fluid’, which is the analogue of the critical background in steady state. As $t$ increases, the pattern is repeated on a larger scale $\mathcal{L}(t)$, leading to the observed property of scaling. 

\section{Distribution of Local Extremum Mass}

Let us investigate the formation of local condensates within regions of size $\mathcal{L}(t)$. The scaled correlation plots in Fig. \ref{correlation} give evidence of a growing length scale $\mathcal{L}(t)=kt^{1/2}$ for both the asymmetric ZRP and the CMAM. In our numerical simulation, we divide the full system of size $L$ into non-overlapping boxes of size $\mathcal{L}(t)$ with $k=1$, ensuring that $\mathcal{L}(t) << L$, to avoid finite size effects. At $t=0$, we place $M = \rho\,L$ particles randomly on the $L$ sites, with $\rho > \rho_c$. The system is evolved using random sequential update. At time $t$, there is a set of $L/\mathcal{L}(t)$ boxes, and we determine the largest occupancy $m^*$ in each box, i.e. $m^*={\rm max}\{m_i\}$ with $i =1, 2…,\mathcal{L}(t)$. The probability distribution $P(m^*,t)$ is built up by sampling every box at a certain time $t$, and repeating the procedure with distinct random initial conditions.

\begin{figure}[H]
    \includegraphics[width=7.5cm, height=4.5cm]{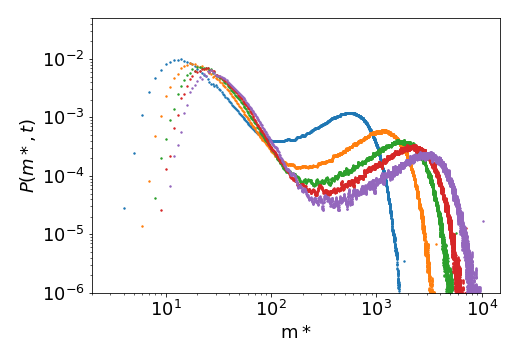}\\
    \includegraphics[width=7.7cm, height=4.6cm]{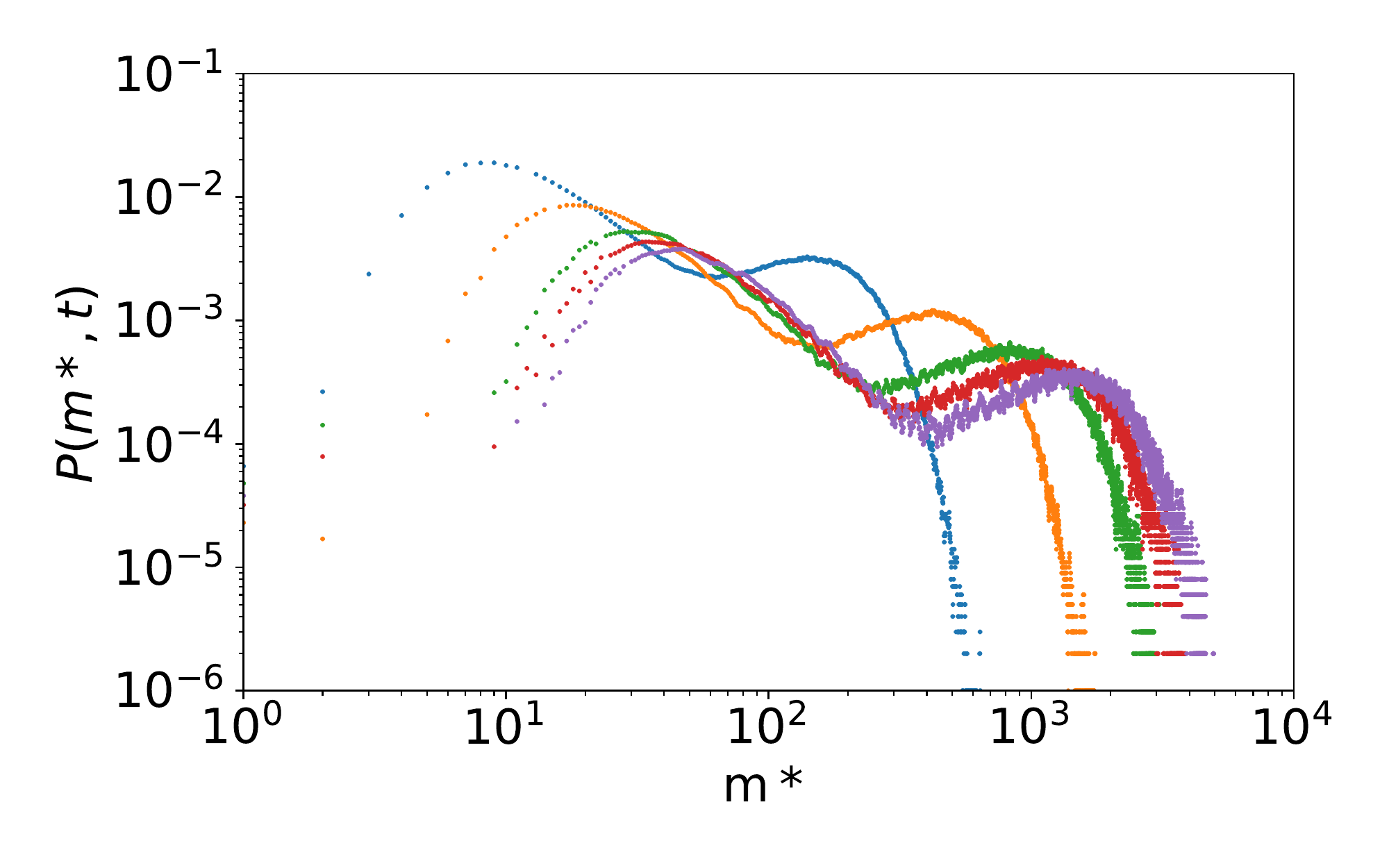}
 \caption{ \it The probability distribution of the maxima in boxes of size $\mathcal{L}(t), P\left(m^{*}, t\right)$ in the two models. Top panel: Asymmetric ZRP ; L=20000, $\mathcal{L} (t)=400,800,1200,1500,2000$; Bottom Panel: CMAM; $L=20000, \mathcal{L}(t)= 200,600,1200,1600,2000$.
 }\label{pmax-unscaled}
\end{figure}

The distribution of local maximum mass $P\left(m^{*}, t\right)$ at different times are shown in Fig. \ref{pmax-unscaled}. \textcolor{black}{Note that this is a bimodal distribution.}

\begin{figure}[h]
    \includegraphics[width=7.5cm, height=4.5cm]{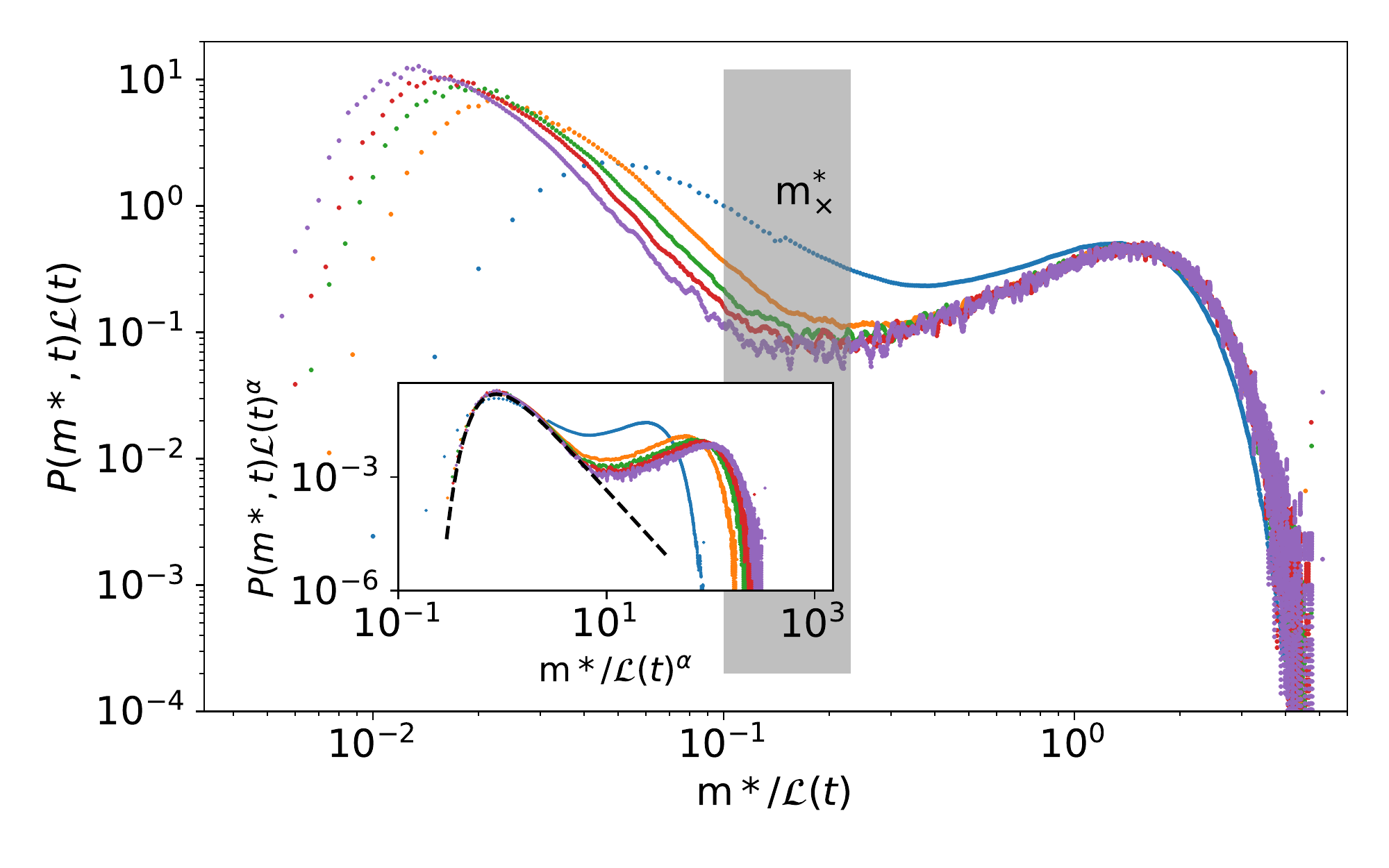}\\
    \includegraphics[width=7.6cm, height=4.6cm]{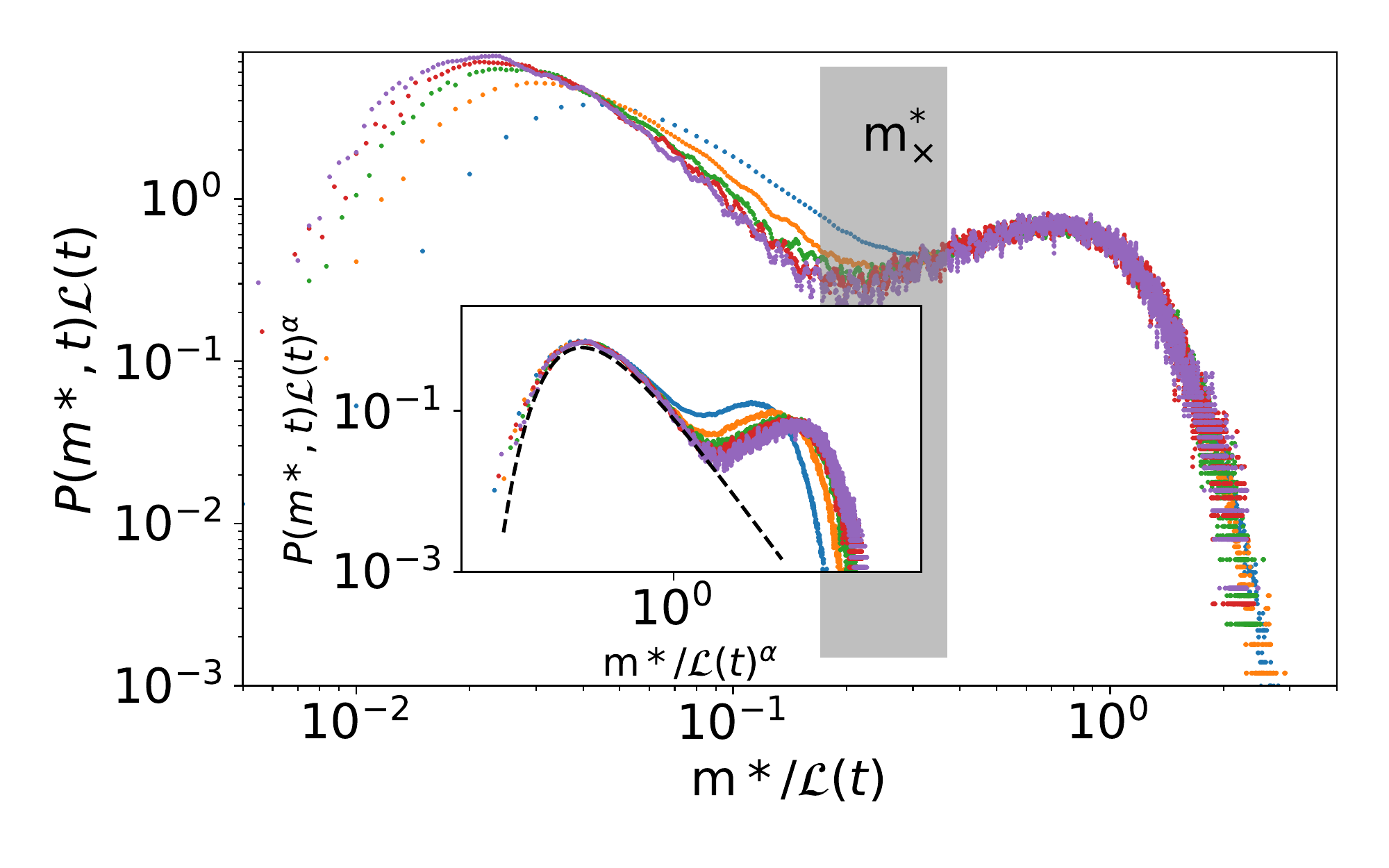}
    
\caption{\it \raggedright Scaled distribution of the largest mass in domains of size $\mathcal{L}(t)$. Top panel: Asymmetric ZRP; Bottom panel: CMAM. Model parameter values as in Fig. \ref{correlation}.
$\mathcal{L}(t)=200, 600, 1200, 1500, 2000$ and system size $L=20 000$. Gray: Crossover region, $(m^* \approx m^*_{\times})$. The region $(m^* > m^*_{\times})$ describes the condensate. For $(m^* < m^*_{\times})$ the plots describe the coarsening critical fluid. \label{local_max}. Insets: Scaled plots for $(m^* < m^*_{\times})$; the dashed line is a
Fr\'{e}chet distribution.}  
\end{figure}

\textcolor{black}{In Fig. \ref{local_max}, we find two regimes in $P\left(m^{*}, t\right)$ exhibiting different scaling behaviour with $\mathcal{L}(t)$ on the left and right of the shaded region $m^*\approx m^*_{\times}$, where $m^*_{\times}$ is a crossover value. This prompts us to enquire whether the full distribution of the box maximum is given by a simple additive scaling form,}

\begin{multline}
P\left(m^{*}, t\right) = c P_{<}+(1-c) P_{>}  \\
= \mathcal{L}(t)^{-\alpha} p_{<}\left(m^{*} / \mathcal{L}(t)^{\alpha}\right)  \\ + \mathcal{L}^{-1}(t) p_{>}\left(m^{*} / \mathcal{L}(t)\right)  \label{scaling-form}    
\end{multline}

$P_{>}(m^{*}, t)$, describes the growing condensates within the $\mathcal{L}(t)$-boxes on coarsening timescales for both the ZRP and CMAM i.e. $m^* > m^*_{\times}$. On the other hand, the region described by $P_{<}(m^{*}, t)$ corresponds to those boxes where there are no large aggregates, i.e. $m^* < m^*_{\times}$. Here, $c$ is the fraction of the area under the $P\left(m^{*}\right)$ curve within the region $\left(m^{*}<m_{\times}^{*}\right)$. We now discuss the properties of $m^*$ in the two regimes.

\subsection{\textbf{$P_{>}(m^{*}, t)$}}

For $m^*  > m^*_{\times}$, Fig. \ref{local_max} shows that the distribution is a scaling function of $m^*/\mathcal{L}(t)$,
\begin{equation}
 P(m^*,t) \sim \frac{1}{\mathcal{L}(t)} p_>\left(\frac{m^*}{\mathcal{L}(t)}\right),~(m^*  > m^*_{\times}). \label{P_R}
\end{equation}
The function $p_>(u)$ is close to Gaussian for ZRP, while it falls exponentially for large $u$ for CMAM. 

For masses greater than $m^*_{\times}$, we identify $m^*$ with $m_{\rm cond}(t)$, the mass of the growing local condensate, which has a value of order $\mathcal{L}(t)$. The fact that $P$ is a scaling function of $m_{\rm cond}(t)/\mathcal{L}(t)$ in the regime $\rho > \rho_c$, implies that the mean $\mu^*_c(t) = \langle m_{\rm cond}(t)\rangle$ and the standard deviation $\sigma^*_c (t) = \sqrt{[\langle m_{\rm cond}(t) ^2\rangle - \langle m_{\rm cond}(t)\rangle^2]}$ are both proportional to $\mathcal{L}(t)$ for both models. The result $\sigma^*_c (t)/\mu^*_c (t) \rightarrow {\rm constant}$ indicates that during coarsening, the behaviour of the condensate mass is {\it fluctuation-dominated}. By contrast, in steady state, $\mu_{\rm ss}=\langle M_{\rm cond}\rangle$ is proportional to $L$ while $\sigma_{\rm cond} \sim L^{\beta}$, with $\beta =1/2$ for ZRP and $\simeq 0.7$ for CMAM, implying $\sigma_{\rm cond}/M_{\rm cond} \rightarrow 0$ for large $L$ indicating relatively mild fluctuations. 
During coarsening, fluctuations of $m_{\rm cond}(t)$ lead to equally strong fluctuations of the total mass within $\mathcal{L}(t)$ (Appendix A). 

\subsection{\textbf{$P_{<}(m^{*}, t)$}}

For $m^*  < m^*_{\times}$, as the locations of the growing condensates are random, several boxes of size $\mathcal{L}(t)$ do not contain a local condensate. Values of $m^* < m^*_{\times}$ correspond to the largest mass in boxes which contain only the coarsening critical fluid, and no condensate.

\begin{figure}[H]
 \begin{center}
 \includegraphics[scale=0.4]{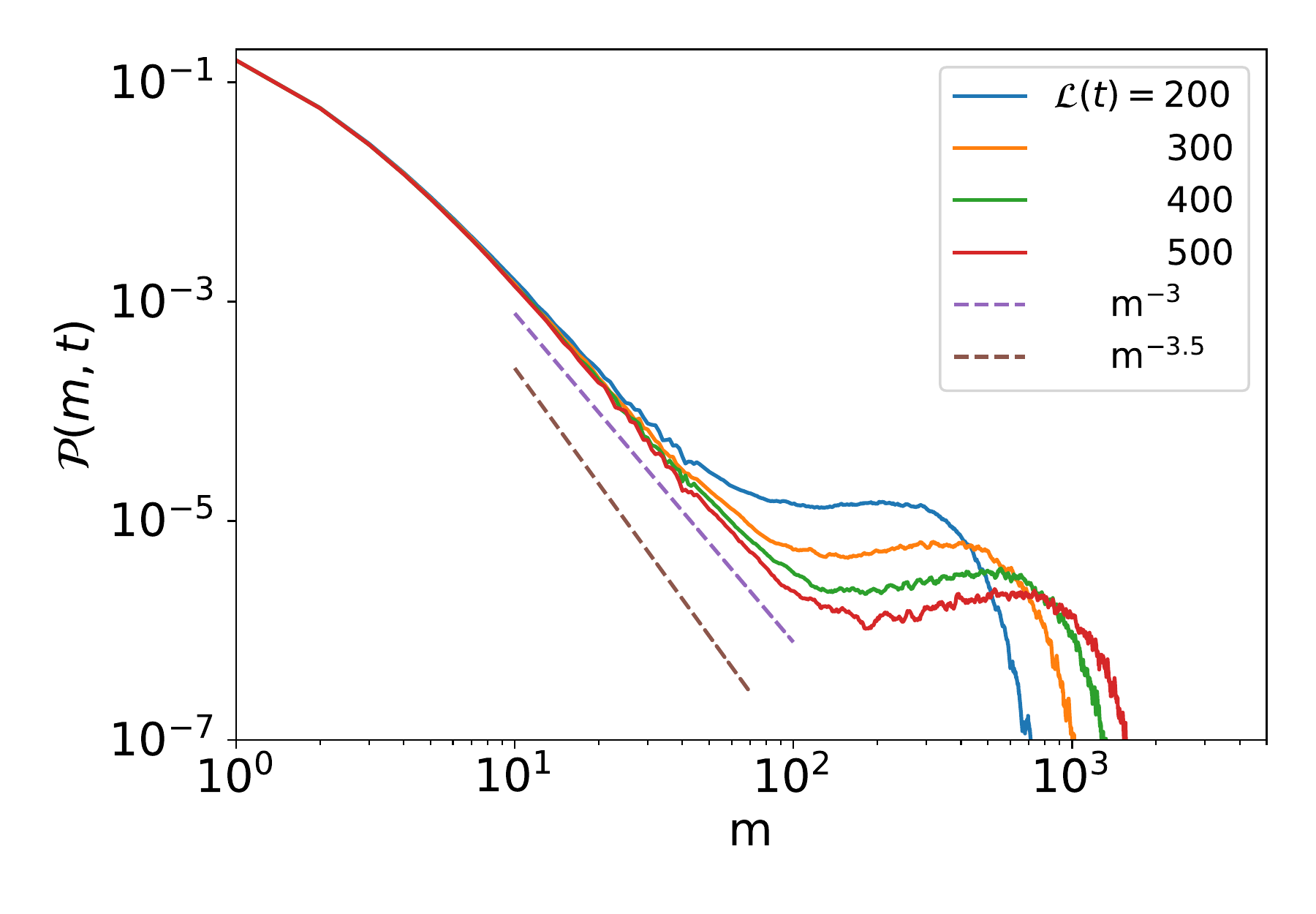}
 \includegraphics[scale=0.27]{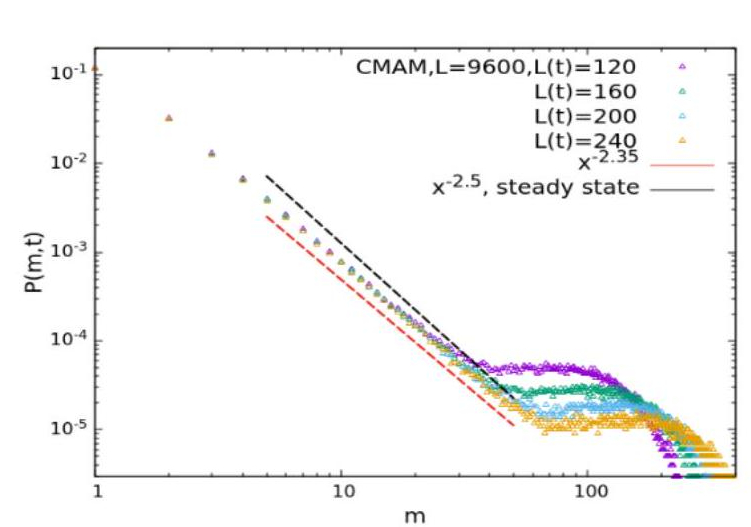}
 \caption{\it The mass distribution $\mathcal{P}(m, t)$ for the $\mathrm{ZRP}$ (left) and the CMAM (right) for various $\mathcal{L}(t)$. For $\mathrm{ZRP}$, the system size $L=10000$, and for CMAM, $L=9600$.}\label{pmfull}
\end{center}
\end{figure}
\vspace{-0.5em}

In this region $m^*$ scales with $\mathcal{L}(t)^{\alpha}$ with $\alpha<1$. Interestingly, the distribution of the scaled variable $u=m^*/\mathcal{L}(t)^{\alpha}$ is proportional to the Fr\'{e}chet distribution $f(u)$ \cite{Tippett} with a power law tail $\sim u^{-\phi}.$ 

\begin{eqnarray}
&~& P_{<}\left(m^{*}, t\right)=\frac{1}{\mathcal{L}(t)^{\alpha}} f(u),\label{pless} \\
 &~& {\rm with}~f(u)=(\phi-1) u^{-\phi}\, e^{-u^{-(\phi-1)}}, u=\frac{m^{*}}{\mathcal{L}(t)^{\alpha}}. 
\nonumber
\end{eqnarray}

Here ($\alpha,\phi$) $\simeq (0.45,3)$ for ZRP and $\simeq(0.7,2.35)$ for CMAM. Note that the area $c$ under this portion of the distribution remains finite even as $\mathcal{L}(t) \rightarrow \infty$, with $c\simeq 0.2$ for ZRP and $\simeq 0.37$ for CMAM.

\textcolor{black}{We rationalize Eq. \eqref{pless} as follows: If we assume the different boxes of size $\mathcal{L}(t)$ carrying the fluid only (sans any condensate) are statistically almost independent, the distribution $P_{<}$ of the maxima in such boxes should be proportional to the extremal distribution for power-law distributed iid variables, namely a Fréchet distribution with shape parameter $\phi$ and scale $\mathcal{L}(t)^{\alpha}.$ The evidence for this argument comes from the {\it mass distribution} $\mathcal{P}(m, t)$ (as opposed to the local maximum mass $m^{*}$) shown in Fig. \ref{pmfull}.
The power law decay exponent in $\mathcal{P}(m, t)$ is identical to the exponent $\phi$ in the Frechet form (Eq. 7).}

Note that the observed values of the power law decay exponent $(\phi \simeq 3$ for the ZRP and $\simeq 2.35$ for CMAM $)$ are smaller than the corresponding steady state values $\left(\phi_{\mathrm{ss}}=3.5\right.$ and $2.5$ respectively), which furthers our claim that the steady-state correspondence during coarsening breaks down in the aggregation-fragmentation models considered herein. 

\begin{figure}[h]
 \begin{center}
 \includegraphics[scale=0.35]{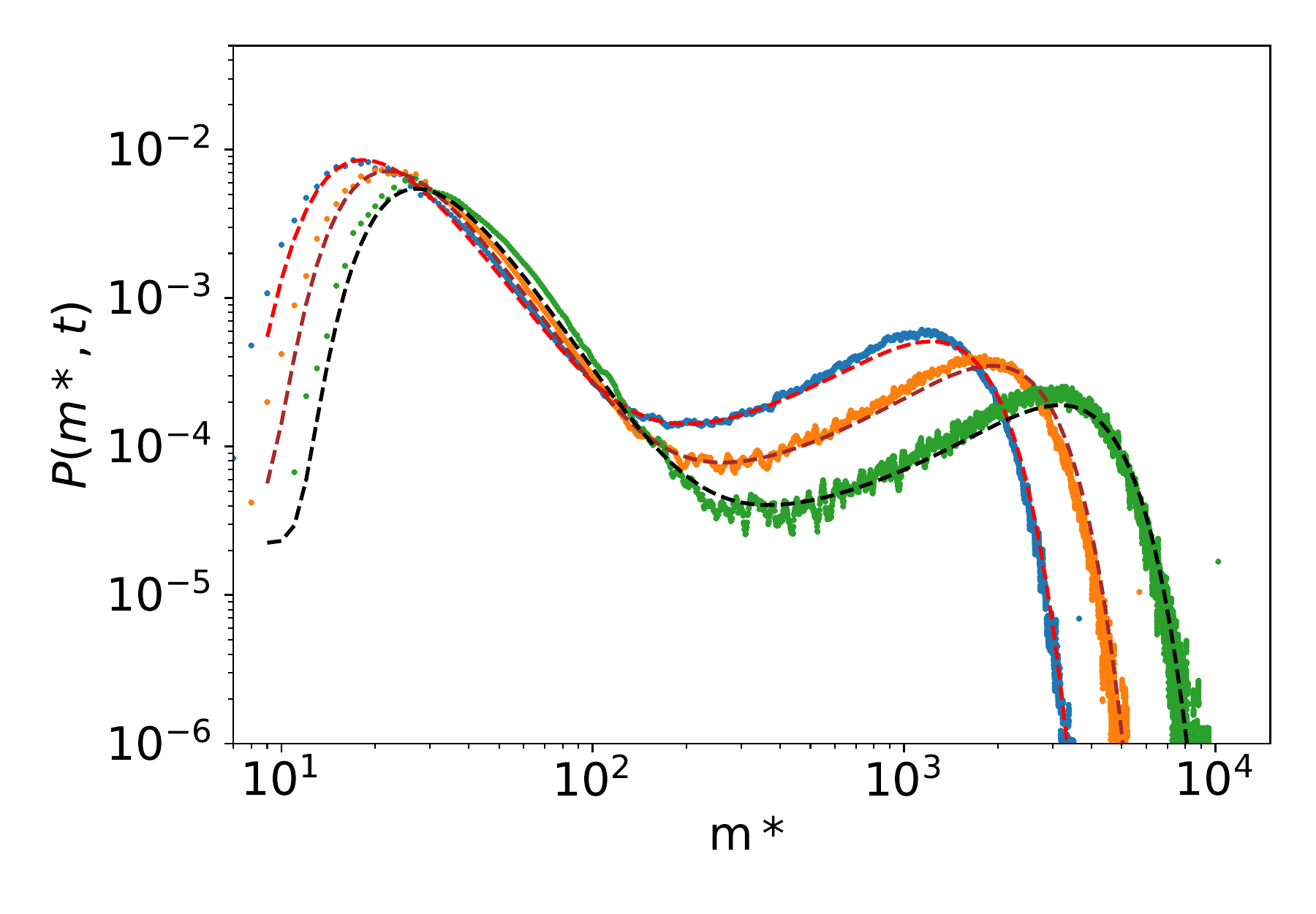}
 \caption{\it \raggedright Distribution of the local maximum mass, $P\left(m^{*}, t\right)$ for $\mathrm{ZRP}$ with $\mathcal{L}(t)=800,1200,2000$, and $L=20000$. This is consistent with the additive scaling form in Eq. \ref{scaling-form} (shown with dashed line).}\label{fit}
\end{center}
\end{figure}

\textcolor{black}{The scaling function obtained using Eqs. \eqref{P_R} and \eqref{pless} in Eq. (5) fits consistently with the data for $P(m^*,t)$, as shown in Fig. \ref{fit}.}

\section{Global Maximum Mass}

\begin{figure}[H]
    \includegraphics[width=7.5cm, height=4cm]{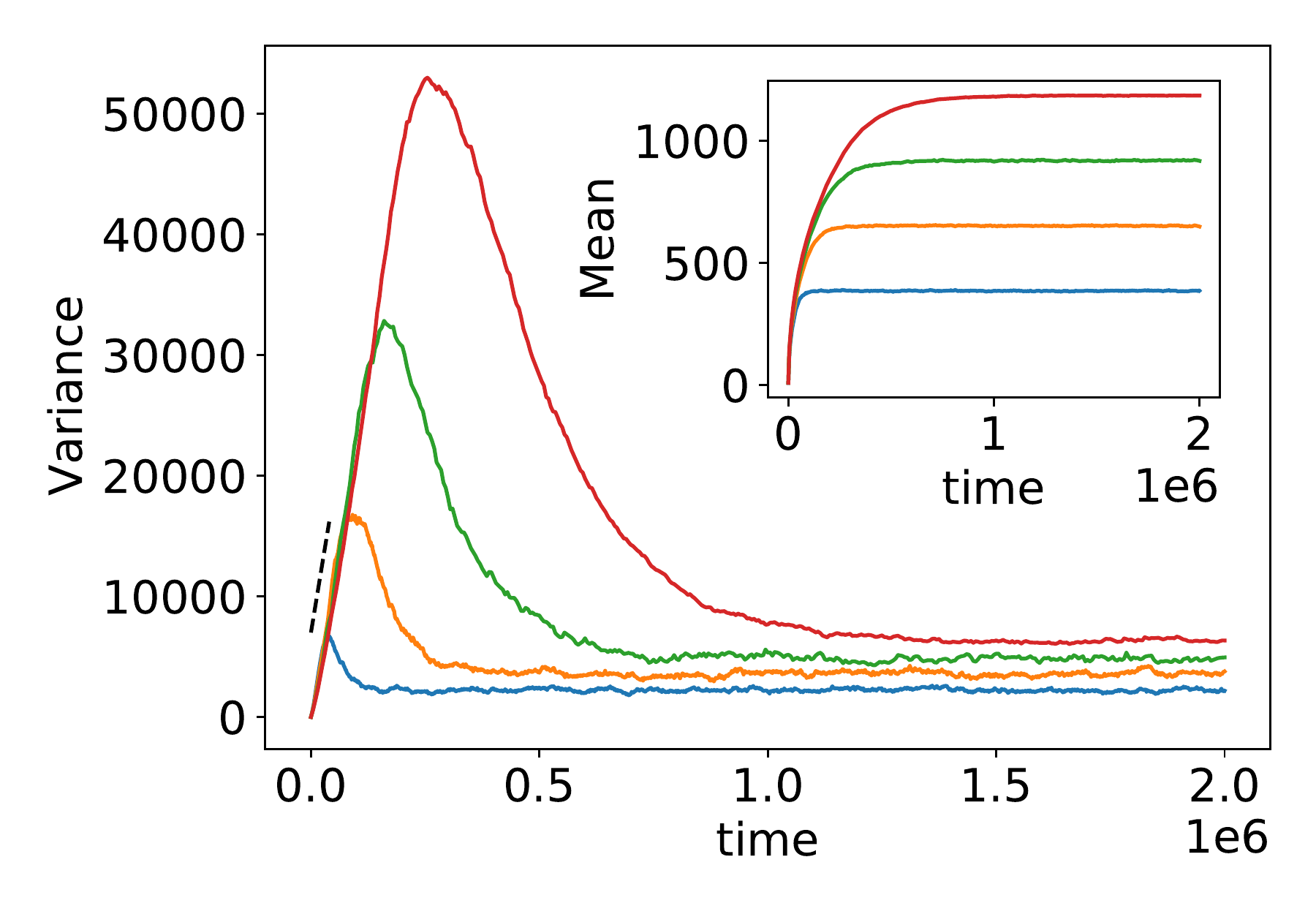}
    \includegraphics[width=7.5cm, height=4cm]{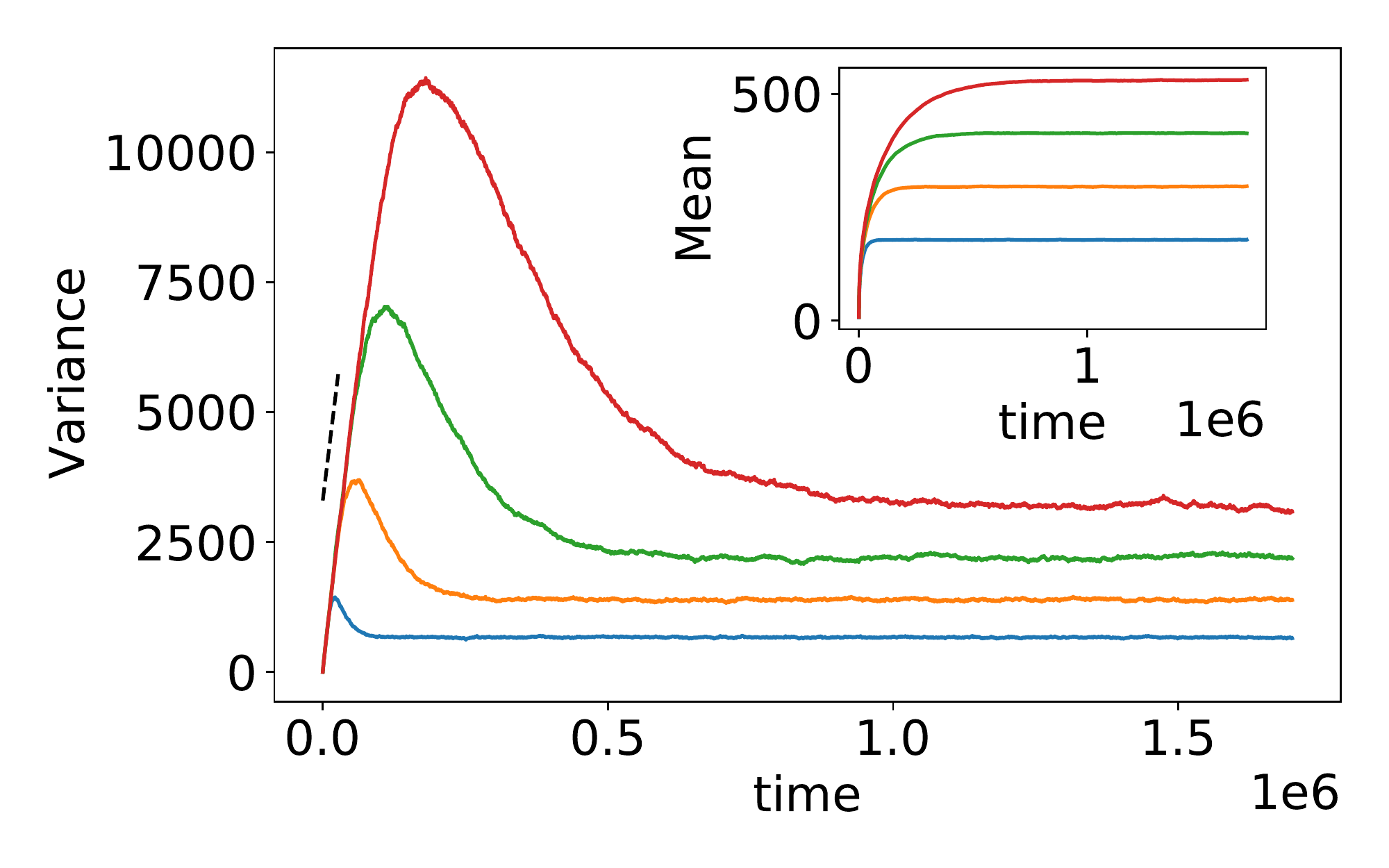}
\caption{\it Variance of the global maximum mass with different system sizes $L=300,500,700,900$. Top panel: ZRP; Bottom panel: CMAM. Model parameter values as in Fig. \ref{correlation}. Insets show that the mean increases smoothly to its steady state value $\sim \mathcal{O}(L)$. The variance increases rapidly and overshoots to a value $\sim \mathcal{O}(L^2)$, finally settling to the much smaller steady state value $\sim \mathcal{O}(L^{2\beta})$, with $\beta=1/2$ for ZRP and $\simeq 0.75$ for CMAM. Note the discrepancy between fluctuations during coarsening and those in steady state.}\label{global_max}
\end{figure}

If $L$ is large but finite, the steady state is reached when $\mathcal{L}(t) \rightarrow L$. The coarsening regime with several local condensates evolves to a regime with a few condensates, and ultimately a single condensate. How does $\sigma^*_c(t)\sim \mathcal{L}(t)$ evolve into $\sigma_{\rm cond}\sim L^{\beta}$ with $\beta<1$ as $\mathcal{L}(t)\rightarrow L$? To shed light on this we study the time evolution of the global maximum mass $M^*$, i.e. the mass on the site with largest occupancy in the system. Fig. \ref{global_max} shows the time evolution of the mean and variance of $M^*$ for both ZRP and CMAM. 

While the mean evolves monotonically (inset of Fig. \ref{global_max}), the variance is strongly nonmonotonic: An initial steep growth in the coarsening regime leads to an overshoot at large times, the maximum being $\mathcal{O}(L^2)$. This is followed by a relaxation regime with the variance approaching the asymptotic steady state value $\sim L^{2\beta}$.

\subsection{Pre-Asymptotic State}

\textcolor{black}{Examination of the state near the peak of the variance reveals that the configurations have a few large condensates. The kinetics in this region is determined by the mass exchange among the large condensates in the ZRP \cite{Grossinsky2003,Evans2005} and by their coalescence in the CMAM. This feature holds also for ZRP with symmetric hopping (Appendix C), and for both mass models in two dimensions as discussed later in the section.}

At some instant, let the masses held by the condensates be $xM_{\rm cond}$ and $(1-x)M_{\rm cond}$, where $M_{\rm cond}\approx (\rho - \rho_c)L$. If $x<1/2$, the largest mass is $M^*=(1-x)M_{\rm cond}$. At some time $\mathcal{O}(L^2)$ later, let the condensate masses be $yM_{\rm cond}$ and $(1-y)M_{\rm cond}$ implying a transfer of mass of $\mathcal{O}(L)$. The change $\Delta M^*$ in $M^*$ is given by $(x-y)M_{\rm cond}$ if $y< 1/2$, or $(x+y-1)M_{\rm cond}$ if $y> 1/2$. Consequently the standard deviation of $\Delta M^*$ is also of order $M_{\rm cond}$.
\textcolor{black}{The mechanism mentioned above leading to $\mathcal{O}(L)$ fluctuations in the 2-condensate pre-asymptotic regime is responsible for the $\mathcal{O}(\mathcal{L}(t))$ fluctuations in the coarsening regime as well.}

\subsection{Scaling of the Global Maximum Mass}

\begin{figure}[h]
 
 \includegraphics[scale=0.4]{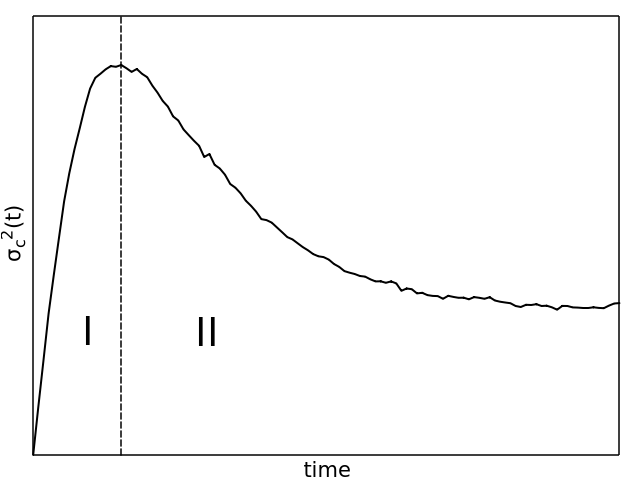}
 \caption{\it \raggedright The time evolution of the variance of the global maximum mass in ZRP and CMAM shows distinct regimes: the coarsening regime (I) and relaxation regime (II).}\label{regimes-schematic}

\end{figure}

\textcolor{black}{The variance $\sigma_{c}^{2}(t, L)$ shows striking nonmonotonic behaviour. There are two distinct regimes (Fig. \ref{regimes-schematic}): (I) a sharply increasing coarsening regime, (II) a relaxation regime where the variance falls and eventually reaches its steady state value as $t \rightarrow \infty$.}

\begin{figure}[H]
 \begin{center}
 \includegraphics[scale=0.14]{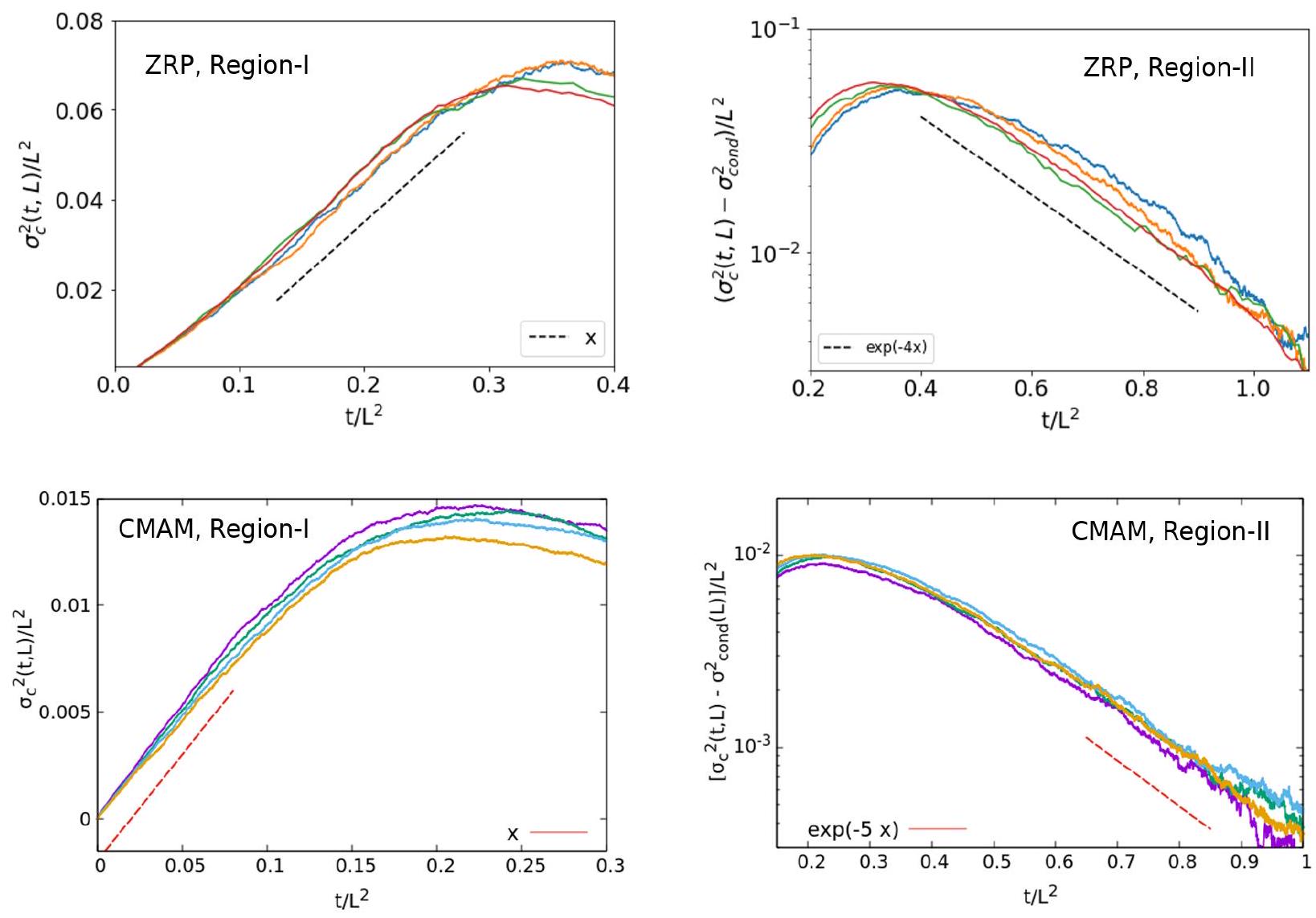}
 \caption{\it The scaling analysis of the variance of the global maximum mass for different system sizes. The top panels are for ZRP $(L=440,512,700,900)$ and bottom panels are for CMAM $(L=500,700,900,1200)$. The coarsening and relaxation regimes satisfy Eqs. \eqref{var-left}-\eqref{var-right}.} \label{regions-fig}
\end{center}
\end{figure}

\textcolor{black}{We find that for both ZRP and CMAM, the variance is well described by separate scaling functions of $u=t / L^{2}$ in the two regions,}

\begin{eqnarray}
&& {\rm Region-I :} \sigma_{c}^{2}(t, L)=L^{2} g_{1}\left(t / L^{2}\right), \nonumber \\  &&~~~{\rm with}~ g_{1}(u) \sim u ~{\rm for ~small}~ u, \label{var-left}  \\
&& {\rm Region-II :} \left[\sigma_{c}^{2}(t, L)-\sigma_{\rm cond}^{2}(L)\right]=L^{2} g_{2}\left(t / L^{2}\right), \nonumber\\ &&~~~{\rm with} ~g_{2}(u) \sim \exp (-a u) ~ {\rm for ~large}~ u. \label{var-right}
\end{eqnarray}
~\\
\textcolor{black}{Here $\sigma_{\text {cond }}^{2}(L) \propto L^{2 \beta}$ is the variance in the steady state, with $\beta=0.5$ for ZRP and $\simeq 0.7$ for CMAM. The data shown in Fig. \ref{regions-fig} seems to converge to Eqs. \eqref{var-left}-\eqref{var-right} as $L$ increases.\\}

\textcolor{black}{It is instructive to look at typical configurations in the different regimes. A typical configuration in region-I consists of several condensates whose number decreases as $t$ increases, finally reaching a state with two condensates. These merge into a single condensate around the time when the variance becomes maximum, after which (region-II) the mass of the condensate increases steadily, finally saturating to its steady state value. Thus, regions I and II are associated with different physical effects, namely coarsening and relaxation, respectively.}

\textcolor{black}{Scaling Laws for the Mean of the Global Maximum mass are given in Appendix B.}

\subsection{Fluctuation of Global Maximum Mass in 2 Dimensions}
\begin{figure}[h]
 
 \includegraphics[scale=0.35]{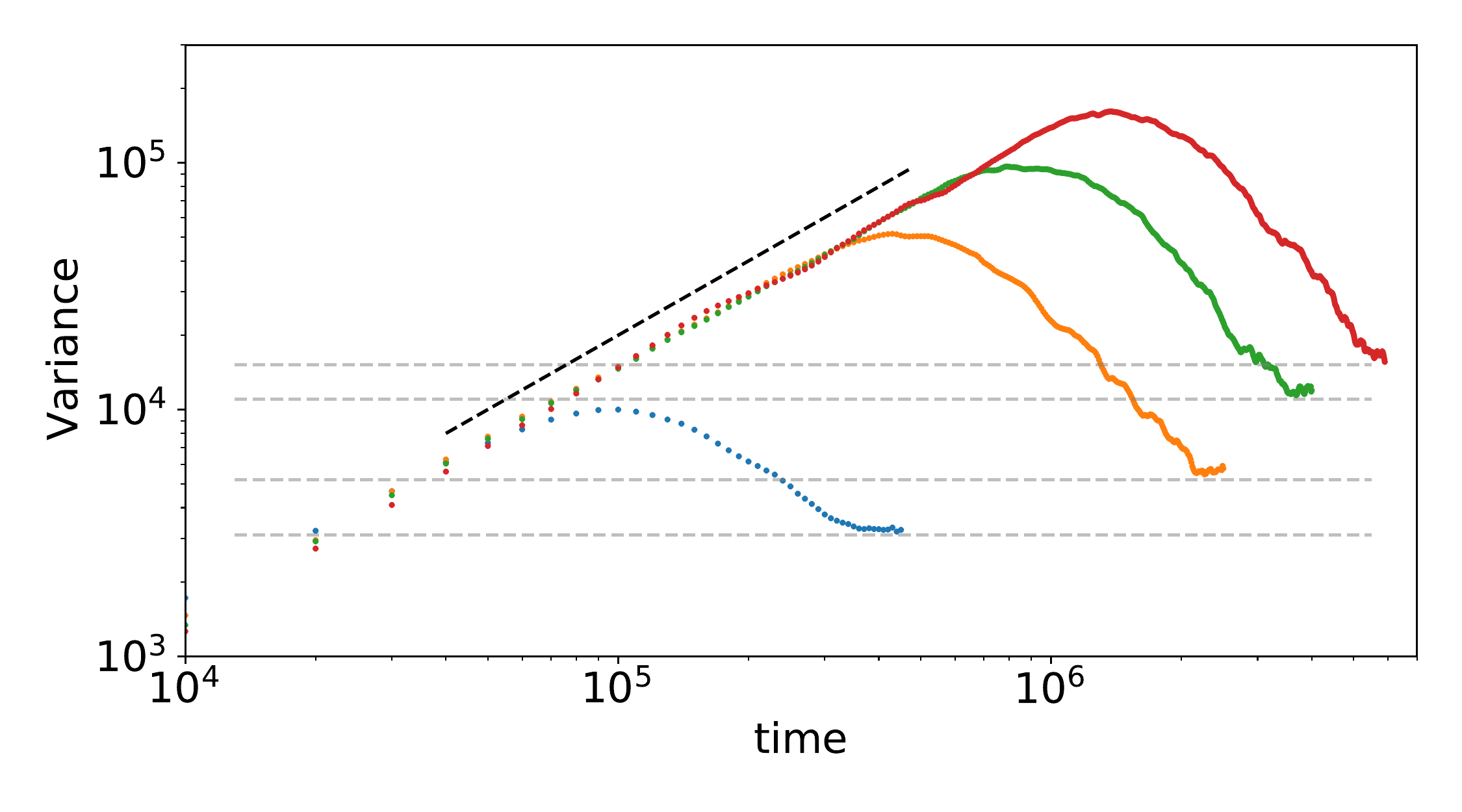}\hspace{0.5em}
 \includegraphics[scale=0.35]{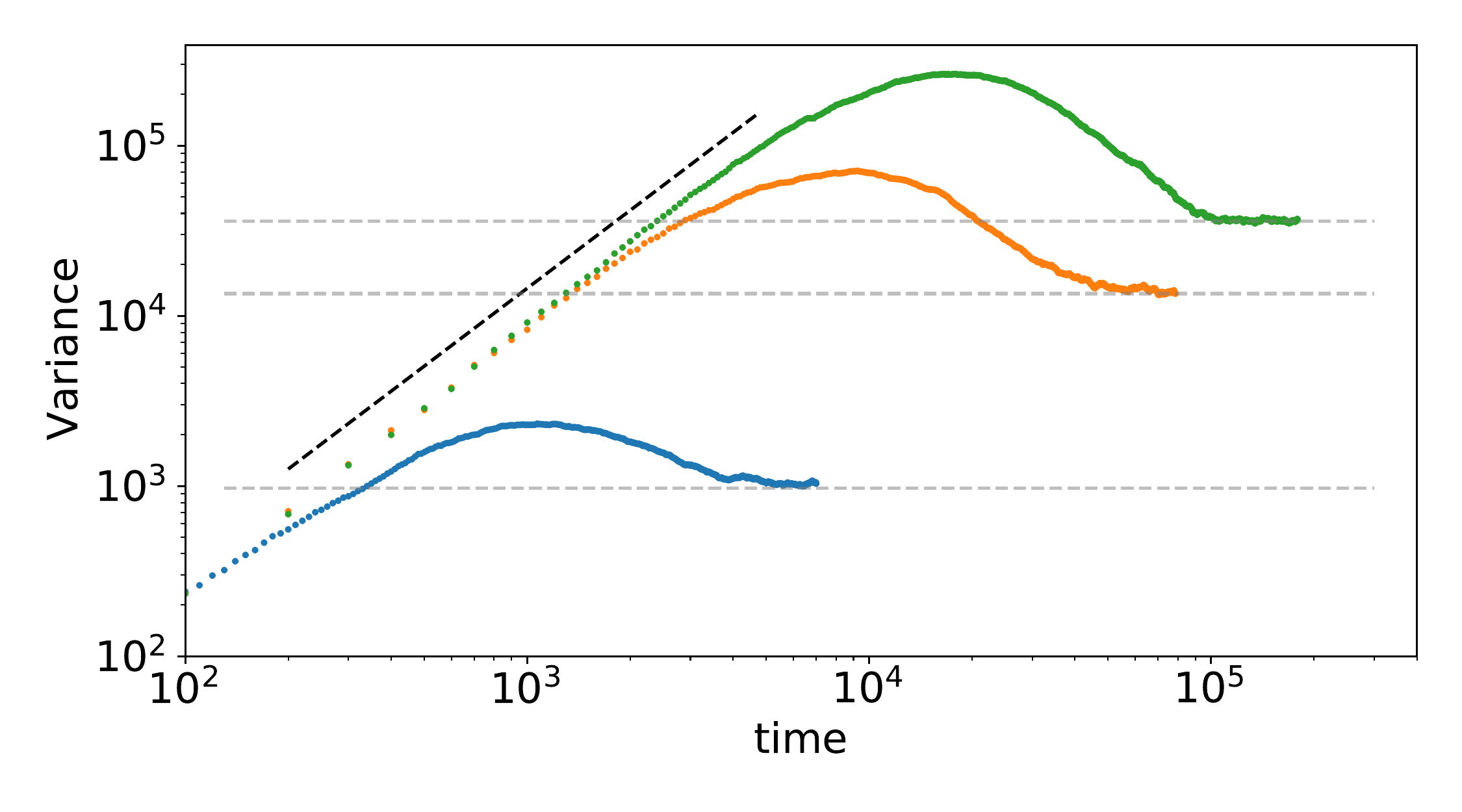}
 \caption{\it \raggedright Variance of the maximum mass in 2D asymmetric $\mathrm{ZRP}$ (top panel) and the CMAM (bottom panel) as a function of time. The data is for system sizes $A=L \times L=20 \times 20,30 \times 30,35 \times$ $35,40 \times 40$ for $\mathrm{ZRP}$ and $A=20 \times 20,50 \times 50,70 \times 70$ for CMAM. The dashed line indicates growth of the variance as $\mathcal{A}(t)^{2}$.}\label{2D}

\end{figure}

\textcolor{black}{In 1D we find that the global maximum mass exhibits anomalously large fluctuations during coarsening. The variance reaches a very high value $\sim O\left(L^{2}\right)$ in the pre-asymptotic regime, before settling down to a value $\sim O\left(L^{2 \beta}\right), \beta<1$ in the steady state. Similar behaviour is observed in $2 \mathrm{D}$ as well. In Fig. \ref{2D} the time evolution of the variance of the global maximum mass is shown for the $2 \mathrm{D}$ asymmetric $\mathrm{ZRP}$ and CMAM in an $L \times L$ system of area $A=L^{2}$. While coarsening, we find that the variance of the largest mass in a domain of typical length $\mathcal{L}(t) \sim t^{1 / z}$, grows as $[\mathcal{L}(t)]^{4} \propto \mathcal{A}(t)^{2}$ in both models, implying giant fluctuations. Our numerical analysis gives $z \simeq 2$ in ZRP and $z \simeq 2.67$ in CMAM. In the steady state, however, the variance of the condensate mass is $\sim O\left(A^{2 \beta}\right)$, where $\beta=0.5$ in $\mathrm{ZRP}$ and in CMAM, we observe $\beta \simeq 0.7$. We find that the peak value of variance in both models is $\approx O\left(A^{2}\right)$. This implies the fluctuation in the largest mass value grows anomalously fast in the coarsening regime even in $2 \mathrm{D}$ and the correspondence to the steady state is lost.}

\section{Exclusion Models}

\begin{figure}[h]
 
 \includegraphics[scale=0.4]{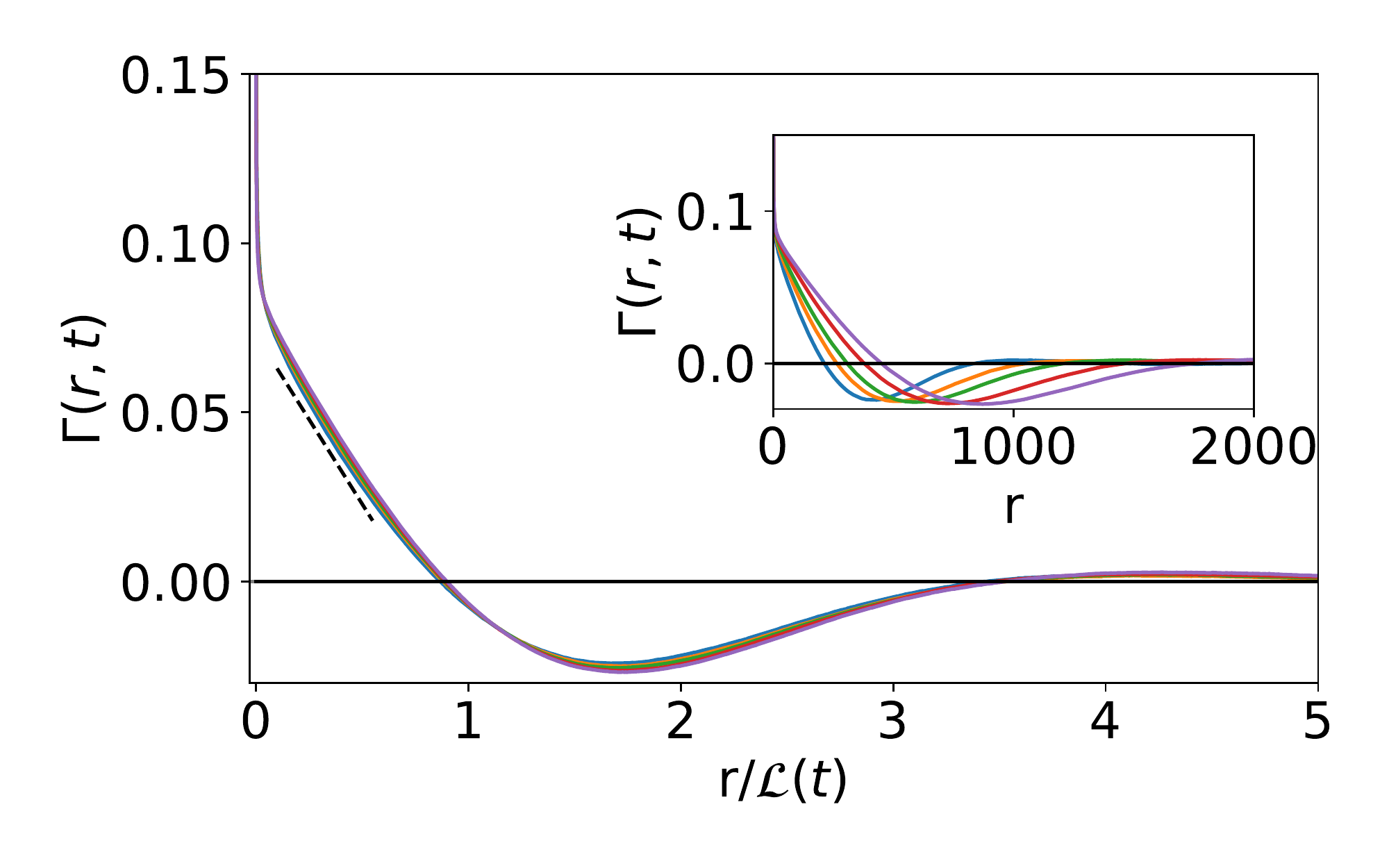}
 \includegraphics[scale=0.4]{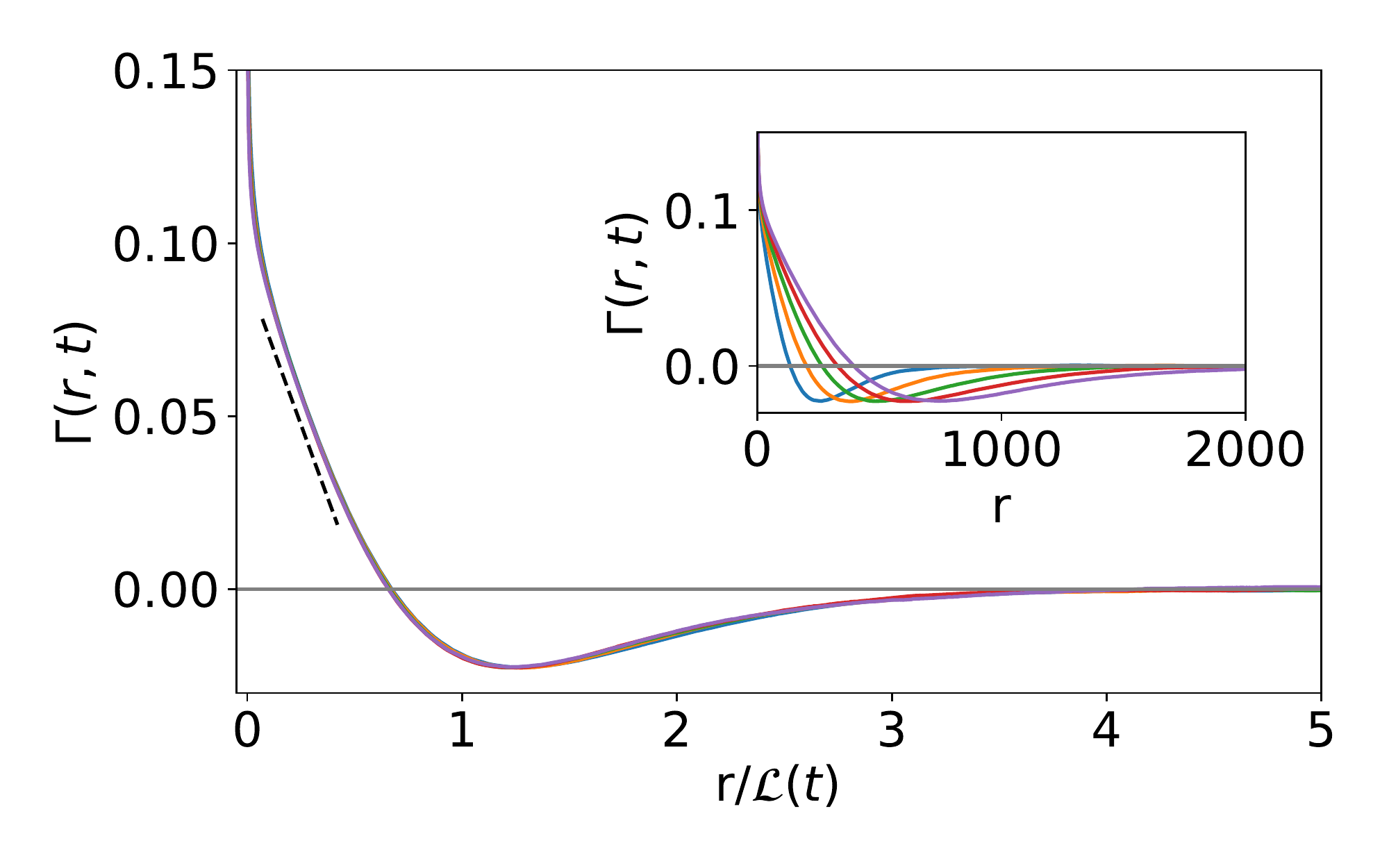}
 \caption{\it \raggedright In the exclusion models, $\Gamma(r, t)$ is a function of the scaled variable $r / \mathcal{L}(t), \mathcal{L}(t)=t^{1 / 2}$. The model parameters are same as in Fig. 2 in the main text. Top panel: EM-ZRP, with $\mathcal{L}(t)=$ $245,300,346,424,500$; Bottom panel: EM-CMAM, with $\mathcal{L}(t)=200,300,400,500,600$. $\Gamma(r, t)$ initially falls linearly, parallel to the dashed lines, indicating the validity of Porod Law. Insets: Unscaled plots.}\label{corr-traffic}

\end{figure}

An exact mapping connects the 1D mass models studied above to exclusion models (EMs): lattice sites in the parent model map to labelled hard-core particles, and particles in the parent model become holes in the EMs \cite{Krug-Ferrari,Evans-Traffic,Evans2005,Kavita-Priyanka,Tanmoy2020}. As illustrated in Fig. \ref{schematic}, particles in the exclusion models move either with short range hops with headway-dependent rates (EM-ZRP), or short and long range hops (EM-CMAM). The occupancy of each site is $\eta_i \in \lbrace 0,1 \rbrace$. The mapping implies that (a) the steady state condensate in the parent model corresponds to phase separation in the EMs, with formation of a cluster with a macroscopic number of holes; (b) during coarsening, the variance of the largest headway shows anomalously large fluctuations. \textcolor{black}{The EMs exhibit fluctuation-dominated phase ordering (FDPO) \cite{Dibyendu2001} while coarsening, although fluctuations in the steady state are quite contained.} The two-point correlation $\Gamma(r,t)=\langle \eta_i(t)\eta_{i+r}(t) \rangle - {\bar \rho}^2$ is a function of scaled separation $r/t^{1/2}$ \textcolor{black}{(Fig. \ref{corr-traffic})}, where ${\bar \rho}=\rho/(1+\rho)$ is the density in the EMs. Unlike other systems which exhibit FDPO \cite{Dibyendu2001}, here the scaling function is consistent with the Porod Law \cite{Porod}.

\section{Conclusion}

Our studies of mass models which support a condensate reveal an interesting new scenario during coarsening. On the scale of the growing coarsening length $\mathcal{L}(t)$, the local condensate peak exhibits scaling. There are anomalously large fluctuations, with both mean and standard deviation scaling with $\mathcal{L}(t)$. The state of the system during coarsening is quite unlike the steady state; rather it resembles a pre-asymptotic state, in which the largest mass fluctuates strongly. We have shown that these features hold also for the ZRP with symmetric hopping (Appendix C) which approaches an equilibrium state, and results for the variance of the global maximum suggest that a similar scenario operates in 2D mass models as well. It would be interesting to see whether fluctuation-dominated coarsening regimes operate in other systems also, for instance with quenched disorder, as in the traffic models studied in \cite{Krug-Ferrari,Evans-Traffic}, or more generally in other driven systems \cite{Baldassari2015,Katyal2020,Levis2014}.

Acknowledgements: C.I. thanks UM-DAE CEBS Mumbai and TIFR Hyderabad for hospitality and academic support. He acknowledges the KVPY fellowship of the Department of Science and Technology (DST), India for financial assistance. M.B. acknowledges support under the DAE Homi Bhabha Chair Professorship of the Department of Atomic Energy, India. This project was funded by intramural funds at TIFR Hyderabad from the Department of Atomic Energy (DAE), India. We acknowledge the useful suggestions of the referee which helped improve the article.

\newpage

\section*{Appendix}

\section*{A. Total mass fluctuations within $\mathcal{L}(t)$} \label{tot-mass}

\begin{figure}[h]
 
 \includegraphics[width=7cm,height=5cm]{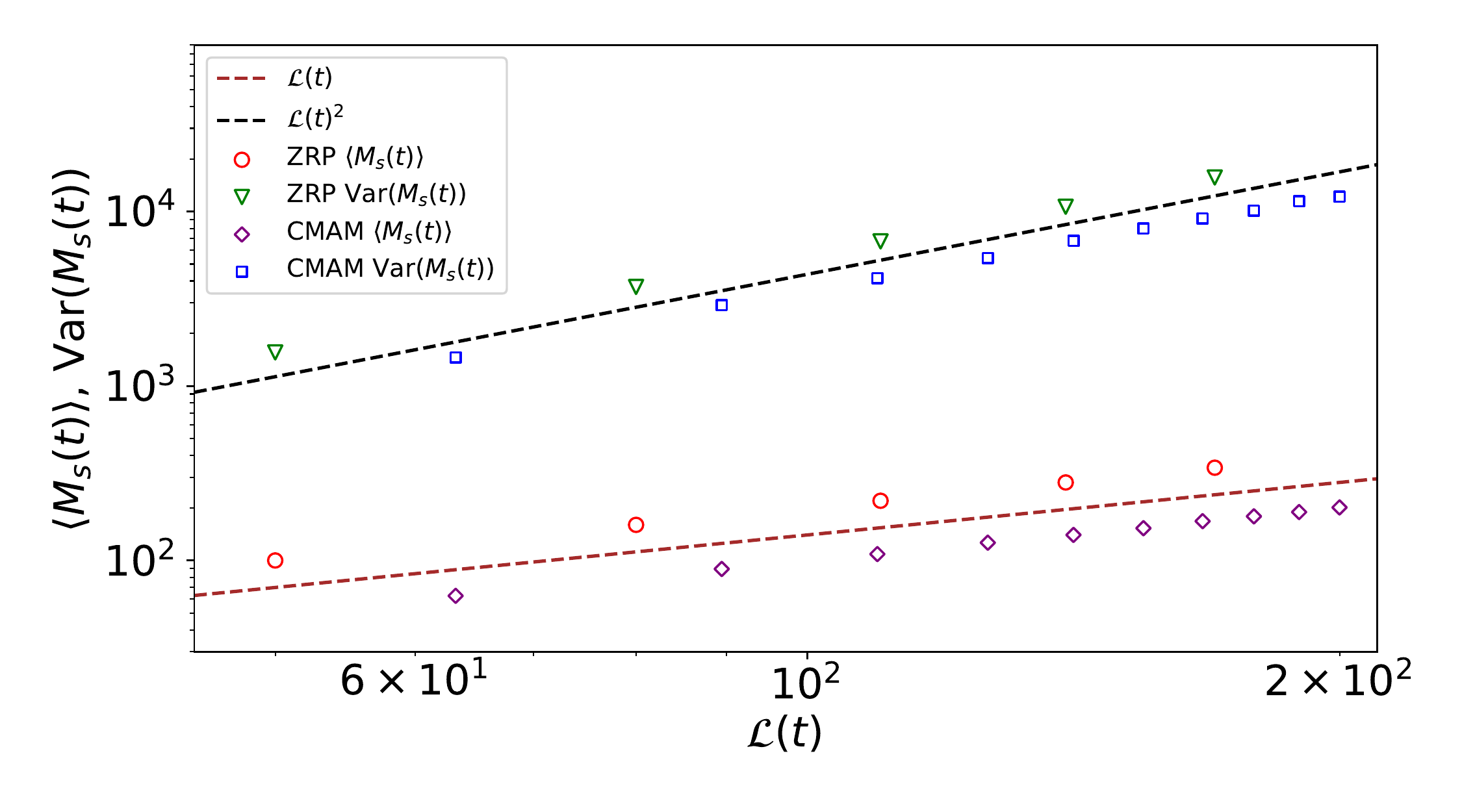}
 \caption{\it \raggedright Mean and variance of the total mass within domains of size $\mathcal{L}(t)$ in the coarsening regime, for $\mathrm{ZRP}$ and CMAM.}\label{totmass}

\end{figure}

\textcolor{black}{The time evolution of the average and variance of the total mass within a domain of size $\mathcal{L}(t)=t^{1 / 2}$ is shown in Fig. \ref{totmass} for CMAM and ZRP. Note that both the mean and standard deviation of the total mass scale as $\mathcal{L}(t)$. This behaviour is similar to that of the largest mass in $\mathcal{L}(t)$ as discussed in the text, indicating that the behaviour of the total mass content in $\mathcal{L}(t)$ follows that of the local condensates. The data shown is for system size $L=1000$ for ZRP and $L=500$ for CMAM.}

\section*{B. Mean of the global maximum mass} \label{mean-max}

\begin{figure}[h]

\includegraphics[scale=0.4]{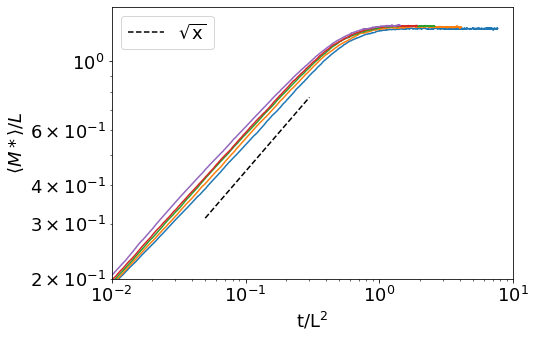}\\
\includegraphics[scale=0.3]{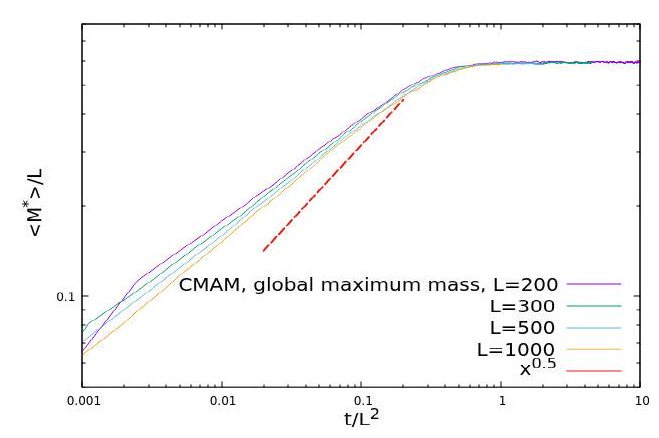}
 \caption{\it \raggedright The scaling analysis of the average global maximum mass for different system sizes. The top panel for the ZRP and bottom panel for the CMAM. The coarsening regime for CMAM shows the lack of scaling and the deviation from the $\sqrt{t}$ behaviour is shown by the dashed lines.} \label{max-mean}
\end{figure}

\textcolor{black}{The mean of the global maximum mass $\left\langle M^{*}(t)\right\rangle$, shown in the insets of Fig. \ref{global_max} in the text, shows a monotonic time dependence. The time evolution of $\left\langle M^{*}(t)\right\rangle$ can also be divided into two regimes. The 'coarsening regime' lasts up to a timescale in which the fluctuation in the global maximum mass also reaches its peak. This is followed by an exponentially decaying 'relaxation regime', in which $\left\langle M^{*}(t)\right\rangle-\left\langle M^{*}(t)\right\rangle_{\mathrm{ss}} \sim \exp \left(-a t / L^{2}\right)$ where $a$ is a constant.\\}

\textcolor{black}{One might have expected $\left\langle M^{*}(t)\right\rangle=L f\left(t / L^{2}\right)$, with $f(x) \sim \sqrt{x}$ for small $x$, so as to recover $\left\langle M^{*}(t)\right\rangle \sim \sqrt{t}$, independent of $L$. Fig. \ref{max-mean} however shows some deviation from this behaviour for ZRP and a stronger deviation for CMAM, which indicates a strong $L$-dependence and nontrivial time dependence. We anticipate that this behaviour of $\left\langle M^{*}(t)\right\rangle$ in the coarsening regime may have its origins in logarithmic corrections coming from extremal statistics, as has been seen in some other systems \cite{cdmodels}. A complete analysis along these lines would also need to account for the correlations.}

\section*{C. Symmetric ZRP: maximum mass in $\mathcal{L}(t)$} \label{symzrp}

\textcolor{black}{The steady state measures for the symmetric and asymmetric ZRP are identical. In the symmetric case, however, the condition of detailed balance holds and the steady state is an equilibrium state. During coarsening, we find that $\mathcal{L}(t) \sim t^{1 / 3}$, as was argued in \cite{Grossinsky2003,Evans2005}.\\}

\begin{figure}
 
 \includegraphics[scale=0.4]{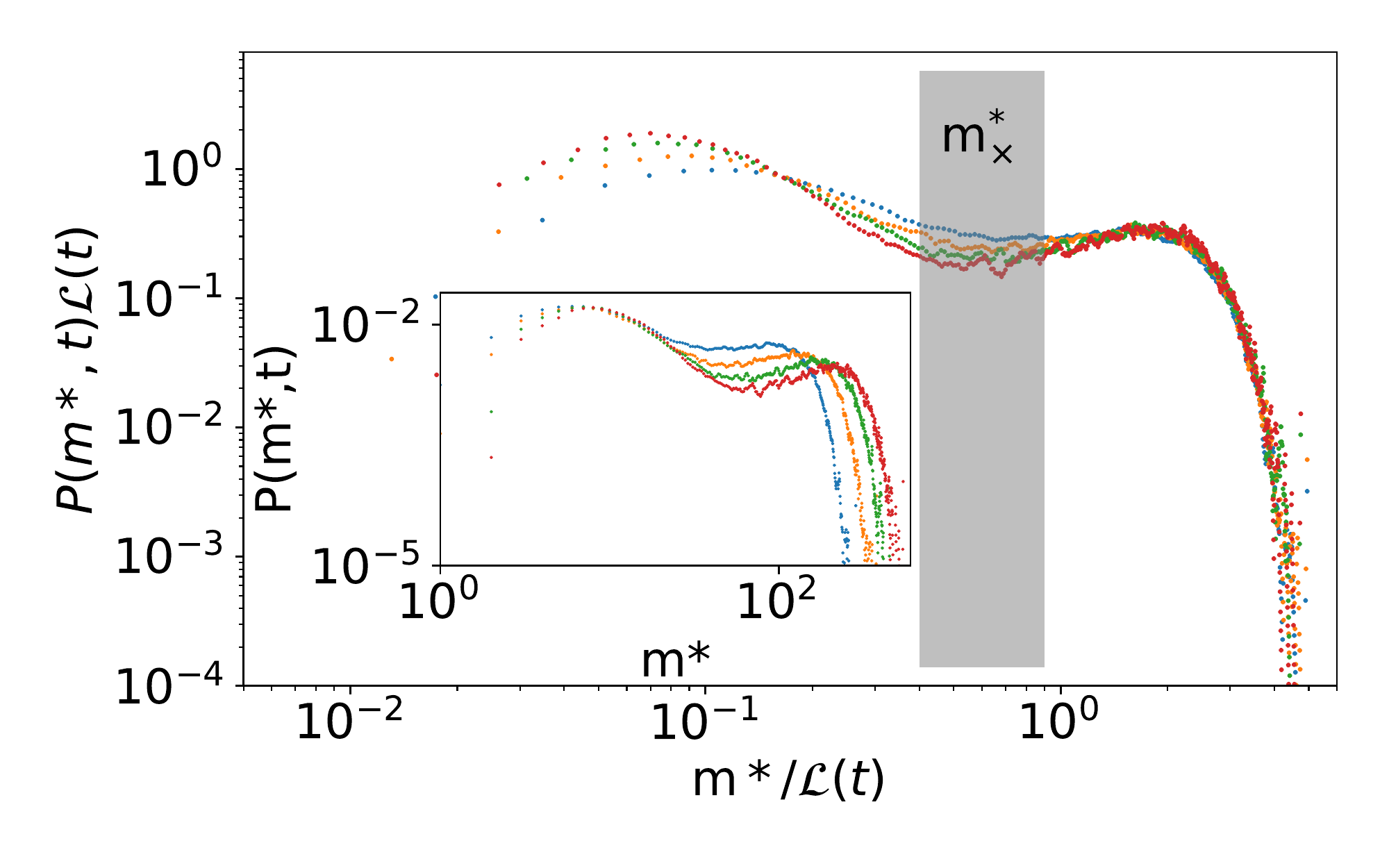}
 \caption{\it \raggedright Distribution and scaling of the maxima in a domain of size $\mathcal{L}(t)$ for the symmetric Zero Range Process in the coarsening regime. The inset shows unscaled plot.}\label{zrp-sym}

\end{figure}

\textcolor{black}{Following Fig. \ref{local_max} in the text for the asymmetric ZRP, we plot the probability distribution function of the local condensates for the symmetric ZRP (Fig. \ref{zrp-sym}). Defining a crossover region $m^{*}=m_{x}^{*}$, we draw the same conclusion: The distribution is bimodal; the region to the right of $m_{\times}^{*}$ a simple scaling function of $m^{*} / \mathcal{L}(t)$, as given by Eq. (5) in the main text. The consequence is: The standard deviation $\sigma_{c}^{*}(t)$ of the fluctuations in largest mass during coarsening is proportional to $\mathcal{L}(t)$. The data (binned) shown is for system size $L=4000$, and the values of $\mathcal{L}(t)$ are $60,80,100$ and 120.}

\end{document}